\documentclass[aps,prd,longbibliography,onecolumn,showpacs,amsmath,amssymb,superscriptaddress,nofootinbib,floatfix,preprintnumbers,secnumarabic]{revtex4-1}

\usepackage[usenames,dvipsnames]{color}
\include{jdefs}
\usepackage{multirow}
\usepackage{graphicx}
\usepackage{adjustbox}
\usepackage{booktabs}
\usepackage{appendix}
\usepackage{slashed}
\usepackage{url}
\usepackage{xspace}
\usepackage{braket}
\usepackage{tabularx}
\definecolor{darkblue}{rgb}{0,0,0.5}
\definecolor{darkgreen}{rgb}{0.1,0,0.3}
\definecolor{darkred}{rgb}{0.6,0,0}
\usepackage[bookmarks=false, bookmarksnumbered=true, colorlinks=true,
  urlcolor=darkblue, citecolor=darkgreen, linkcolor=darkred,
  breaklinks=true]{hyperref} 
\usepackage[capitalise]{cleveref}
\usepackage[normalem]{ulem} 
\usepackage{siunitx}
\usepackage[utf8]{inputenc}
\usepackage{xspace}

\usepackage{fullpage}
\usepackage{geometry}      
\AtBeginDocument{
  \heavyrulewidth=.08em
  \lightrulewidth=.05em
  \cmidrulewidth=.03em
  \belowrulesep=.65ex
  \belowbottomsep=0pt
  \aboverulesep=.4ex
  \abovetopsep=0pt
  \cmidrulesep=\doublerulesep
  \cmidrulekern=.5em
  \defaultaddspace= .5em
  \setlength{\tabcolsep}{0.5em}
}

\newcommand{\ee}{\textsubscript{ee}}
\newcommand{\nr}{\textsubscript{nr}}
\newcommand{\C}[1]{\mathcal{#1}}
\newcommand{\Umt}{$U(1)_{L_\mu-L_\tau}$\xspace}

\newcommand{\coherent}{CE$\nu$NS\xspace}

\begin{document}

\preprint{IPPP/20/24}
\preprint{IFT-UAM/CSIC-20-70}

\vspace*{0.7cm}

\title{
Solar neutrino probes of the muon anomalous magnetic moment in the gauged $U(1)_{L_\mu-L_\tau}$
}

 \author{D.W.P. Amaral}
 \email{dorian.w.praia-do-amaral@dur.ac.uk}
 \affiliation{Institute for Particle Physics Phenomenology, Durham University, Durham DH1 3LE, United Kingdom} 
 \author{D.G. Cerde\~no}
 \email{davidg.cerdeno@gmail.com}
 \affiliation{Instituto de F\' isica Te\'orica, Universidad Aut\'onoma de Madrid, 28049 Madrid, Spain}
 \affiliation{Institute for Particle Physics Phenomenology, Durham University, Durham DH1 3LE, United Kingdom} 
 \author{P. Foldenauer}
 \email{patrick.foldenauer@durham.ac.uk}
 \affiliation{Institute for Particle Physics Phenomenology, Durham University, Durham DH1 3LE, United Kingdom} 
 \author{E. Reid}
 \email{elliott.m.reid@durham.ac.uk}
\affiliation{Institute for Particle Physics Phenomenology, Durham University, Durham DH1 3LE, United Kingdom}

 \date{\today}
 \begin{abstract}

Models of gauged $U(1)_{L_\mu-L_\tau}$ can provide a solution to the long-standing discrepancy between the theoretical prediction for the muon anomalous magnetic moment and its measured value. 
The extra contribution is due to a new light vector mediator, which also helps to alleviate an existing tension in the determination of the Hubble parameter.
In this article, we explore ways to probe this solution via the scattering of solar neutrinos with electrons and nuclei in a range of experiments and considering high and low solar metallicity scenarios.
In particular, we reevaluate Borexino constraints on neutrino-electron scattering, finding them to be more stringent than previously reported, and already excluding a part of the $(g-2)_\mu$ explanation with mediator masses smaller than $2\times10^{-2}$~GeV. 
We then show that future direct dark matter detectors will be able to probe most of the remaining solution. Due to its large exposure, LUX-ZEPLIN will explore regions with mediator masses up to $5\times10^{-2}$~GeV and DARWIN will be able to extend the search beyond $10^{-1}$~GeV, thereby covering most of the area compatible with $(g-2)_\mu$.
For completeness, we have also computed the constraints derived from the recent XENON1T electron recoil search and from the CENNS-10 LAr detector, showing that none of them excludes new areas of the parameter space. 
Should the excess in the muon anomalous magnetic moment be confirmed, our work  suggests  that direct detection experiments could provide crucial  information with which to test the $U(1)_{L_\mu-L_\tau}$ solution, complementary to efforts in neutrino experiments and accelerators.
\end{abstract}

 \maketitle

\section{Introduction}

With the completion of the Standard Model (SM) after the Higgs boson discovery and the absence of any hint for new particles at the energy frontier, the focus of particle physics is shifting towards the exploration of the intensity frontier. At present, increasing emphasis is being put on the emerging discrepancies in precision studies of low-energy observables.
In this context, the anomalous magnetic moment of the muon, $(g-2)_\mu$, stands out as one of the most precisely measured quantities in particle physics. Its determination with the E821 experiment at BNL~\cite{Bennett:2002jb,Bennett:2004pv,Bennett:2006fi} created a now almost two-decade-old puzzle, as the experimental value deviates from its SM prediction by up to $\sim3.7\,\sigma$\footnote{It should be noted that a recent lattice result of the leading order hadronic vacuum polarisation \cite{Borsanyi:2020mff} could significantly reduce this difference, at the expense of worsening fits to other precision EW observables \cite{Crivellin:2020zul}.}~\cite{Keshavarzi:2019abf,Davier:2019can,Aoyama:2020ynm}.
This has been interpreted as a possible signature of physics beyond the Standard Model (BSM), leading to a plethora of constructions with new physics in the leptonic sector~\cite{Appelquist:2001jz,Kim:2001rc,Davoudiasl:2000my,Park:2001uc,Blanke:2007db}.
\par
In an effort to shed new light on the true nature of the observed discrepancy, 
the upcoming measurement at the E989 experiment at Fermilab is projected to determine $(g-2)_\mu$ with a precision of 140 parts per billion~\cite{Holzbauer:2017ntd}. This would result in  a fourfold improvement over the present experimental error~\cite{Grange:2015fou}. Furthermore, an independent measurement of this observable at J-PARC~\cite{Abe:2019thb} with comparable precision to the previous determination at BNL  will be able to  provide further insight on this potential discrepancy.
With these experimental efforts underway and  the first data release from E989 imminent, it seems timely to explore complementary probes of the conjectured explanation of the observed shift in $(g-2)_\mu$. 
\par
In this article, we  focus on a particularly interesting  solution  proposed in terms of a new light mediator associated with a new \Umt gauge symmetry, which received a lot of attention over the last years~\cite{Baek:2001kca,Ma:2001md,Bauer:2018onh,Asai:2018ocx,Chun:2018ibr,Banerjee:2018mnw,Foldenauer:2019dai}. Since gauging the difference between two lepton-flavour numbers is anomaly-free within the SM, these constructions provide a minimal extension of the SM gauge group without the need for extra fermions~\cite{He:1990pn,He:1991qd}. It has been noted that such a model can simultaneously account for the observed  discrepancy in $(g-2)_\mu$ and even accommodate dark matter with the correct relic abundance~\cite{Ma:2001md,Harigaya:2013twa,Altmannshofer:2016jzy,Biswas:2016yjr,Kahn:2018cqs,Foldenauer:2018zrz,Biswas:2019twf}. 
Furthermore, as suggested in Ref.~\cite{Escudero:2019gzq}, a \Umt gauge boson $A'$ 
can significantly alleviate the recent $\gtrsim3\,\sigma$ deviation of local measurements~\cite{Riess:2016jrr,Riess:2018byc,Riess:2019cxk} of the Hubble parameter, $H_0$, from the value inferred from early-time cosmology in  CMB data~\cite{Aghanim:2018eyx}.\footnote{In Ref.~\cite{Blinov:2019gcj} it was, however, pointed out that explanations of the $H_0$ tension by  light vector mediators cannot simultaneously ameliorate the slightly milder tension in the cosmological parameter $\sigma_8$, which is linked to the  small scale power spectrum of the universe.}
In particular, a modification of early-time cosmology with the presence of an effective component of dark radiation of $\Delta N_\text{eff}\sim 0.4$ neutrino-equivalent species could reconcile the two measurements~\cite{Bernal:2016gxb}.

The new \Umt gauge boson does not have couplings to electrons or quarks at leading order.
This makes it hard to test the model in collider or fixed target experiments, which employ electron or hadron beams. However, its tree-level couplings to mu- and tau-flavoured neutrinos open up a unique opportunity to search for such a boson in  processes probing neutrino interactions. Almost all of the dominant current bounds on MeV-mass \Umt bosons result from neutrino probes like neutrino trident production~\cite{Altmannshofer:2014pba}, neutrino-electron scattering in Borexino~\cite{Kaneta:2016uyt}, neutrino cooling of white dwarfs~\cite{Bauer:2018onh} or $N_\text{eff}$ during big bang nucleosynthesis~\cite{Kamada:2015era,Kamada:2018zxi,Escudero:2019gzq}\footnote{In principle, one could also consider constraints from neutrinoless double beta decay induced by the emission of a \Umt gauge boson. However, as the \Umt gauge boson does not couple to first generation neutrinos, one requires at least one additional neutrino mass insertion, changing the neutrino flavor from $e$ to either $\mu$ or $\tau$ (see Sect.~V of ~\cite{Dror:2020fbh} for a detailed explanation). This  makes the process highly suppressed and therefore not relevant in this model.}. The only competitive constraints not relying on neutrinos stem from four-muon searches at BaBar~\cite{TheBABAR:2016rlg} and CMS~\cite{Sirunyan:2018nnz}.

The aim of our work is to explore the sensitivity of solar neutrino scattering off electrons and nuclei as a future independent probe of the region of parameter space relevant for $(g-2)_\mu$. 
We start our analysis by considering previous limits derived from Borexino~\cite{Harnik:2012ni,Bilmis:2015lja} data, and show that these have been generally underestimated. We update these bounds, taking into account the ambiguity associated with the so-called solar metallicity problem, and we perform all computations with solar fluxes as predicted in the solar standard model (SSM) for both the scenario of a low and high metallicity Sun \cite{Vinyoles:2016djt}.

Next, we consider coherent elastic neutrino-nucleus scattering (\coherent). This elusive process  was first detected at the COHERENT experiment in CsI[Na] nuclei~\cite{Akimov:2015nza,Akimov:2017ade}, an observation which has been recently extended to argon nuclei \cite{Akimov:2020pdx}. These results can be used to constrain the parameter space of new light mediators in the neutrino sector and, in fact, the data from the CsI run was used to derive limits for the \Umt model~\cite{Abdullah:2018ykz}. In our work we extend the analysis to derive new bounds from the recent LAr results.

Finally, we investigate the potential of direct dark matter detection experiments to probe the region of the \Umt parameter space consistent with the muon anomalous magnetic moment.
Although originally designed to search for dark matter particles, direct detection experiments can be sensitive to both neutrino-electron elastic scattering and coherent neutrino-nucleus scattering (the latter is usually considered to be an irreducible background for dark matter searches). In the energy range in which we are interested, the main contribution comes from solar neutrinos.
In our analysis we derive constraints from the recent results on electron recoils in XENON1T~\cite{Aprile:2020tmw} and argue that they do not explore new areas of the parameter space of the \Umt model. We then investigate the potential reach of upcoming experiments based on germanium and xenon, inspired by SuperCDMS~\cite{Agnese:2016cpb} and LUX-ZEPLIN (LZ)~\cite{Mount:2017qzi}, as well as third generation detectors inspired on liquid noble gases, such as  DarkSide-20k~\cite{Aalseth:2017fik} and DARWIN~\cite{Aalbers:2016jon}. We find that the optimal strategy to probe the $(g-2)_\mu$ solution is a large exposure, which favours liquid xenon detectors.

This article is organised as follows. In~\cref{sec:model} we introduce the minimal gauged \Umt model, we discuss its impact on the muon anomalous magnetic moment, and we compute the contribution to non-standard neutrino interactions. In~\cref{sec:constraints} we describe the various ways to probe this model via neutrino interactions, in particular via neutrino-electron scattering in Borexino (\cref{sec:constraints_borexino}), via coherent neutrino-nucleus scattering in the COHERENT experiment (\cref{sec:constraints_coherent}), and  via both of these processes in direct dark matter detectors (\cref{sec:constraints_dm}). In~\cref{sec:discussion} we give a detailed discussion of our results,  before we conclude in~\cref{sec:conclusions}.

\section{Minimal gauged $U(1)_{L_\mu-L_\tau}$}
\label{sec:model}

A peculiar feature of the SM Lagrangian is its invariance under the accidental  global symmetries $U(1)_{B,L_e,L_\mu,L_\tau}$. These global symmetries, if  combined into the groups $U(1)_{B-L}$ and $U(1)_{L_i-L_j}$ (with $i,j = e,\mu,\tau$) or linear combinations thereof, can be promoted to anomaly-free gauge symmetries without the addition of any extra new fermion fields\footnote{In case of $U(1)_{B-L}$ one has to add three right-handed neutrinos in order to cancel the anomalies.}. Of these extra anomaly-free  gauge symmetries the case of \Umt is particularly interesting as it still allows for an explanation of the $(g-2)_\mu$ anomaly \cite{Baek:2001kca,Ma:2001md}.

We begin our discussion of models of extra gauged \Umt symmetries by introducing their generic form of the Lagrangian in the gauge basis,
\begin{equation} \label{eq:lagr}
\C{L} = \C{L}_\text{SM} - \frac{1}{4}  X_{\alpha\beta}  X^{\alpha\beta} -\frac{\epsilon_Y}{2}  B_{\alpha\beta}  X^{\alpha\beta} -\frac{M_X^2}{2}  X_\alpha X^\alpha - g_{\mu \tau}\, J^{\mu-\tau}_\alpha X^\alpha \,,
\end{equation}
where $X_\alpha$ denotes the \Umt gauge boson and $B_{\alpha\beta}$ and $X_{\alpha\beta}$ are the field strength tensors of the hypercharge $U(1)_{Y}$ and the new \Umt, respectively.
The parameters $\epsilon_Y$ and $g_{\mu \tau}$ denote the kinetic mixing parameter and gauge coupling constant. The gauge current of the new symmetry in the minimal setup is given by
\begin{equation}
    J^{\mu-\tau}_\alpha = \bar L_2 \gamma_\alpha L_2 
          + \bar \mu_R \gamma_\alpha \mu_R 
          - \bar L_3 \gamma_\alpha L_3 -\bar\tau_R \gamma_\alpha \tau_R\,,
\end{equation}
where $L_2=(\nu_{\mu L}, \mu_L)^T$ and $L_3=(\nu_{\tau L}, \tau_L)^T$ denote the second and third generation $SU(2)_L$ lepton doublets.

In order to study the phenomenology of a gauged \Umt symmetry, we need to determine the interactions of the mass eigenstate $A'_\alpha$ of the associated gauge boson $X_\alpha$.
The kinetic terms in \cref{eq:lagr} are not diagonal and must be diagonalised by a field redefinition of $B_\alpha$ and $X_\alpha$. Subsequent rotation to the mass basis\footnote{For a detailed treatment of diagonalising the kinetic terms and rotating to the mass basis see e.g.~Ref.~\cite{Heeck:2011wj}.} yields interaction terms of the following form~\cite{Bauer:2018onh},
\begin{align}\label{eq:currentcouplings}
{\mathcal{L}}_\mathrm{int}=&-\left(e\, J^\text{EM}_\alpha,\; g_Z\, J^Z_\alpha,\;  g_{\mu\tau}\, J^{\mu-\tau}_\alpha\right) \,K\,\begin{pmatrix}A^\alpha\\ Z^\alpha\\  A^{\prime\alpha}\end{pmatrix} \,, 
\end{align}
with the coupling matrix 
\begin{align}\label{eq:Kmatrix}
K= \begin{pmatrix}
1 & 0 & -\epsilon \phantom{e}\\
0 & 1& 0 \\
0 & \epsilon \tan\theta_W&  1
\end{pmatrix} +\mathcal{O}(\epsilon \delta, \epsilon^{ 2})\,,
\end{align}
expanded to leading order in the small kinetic mixing parameter and   mass ratio of the neutral bosons,
\begin{equation} \label{eq:small_pars}
\epsilon=\epsilon_Y\,\cos\theta_W\,,\qquad\qquad\delta=\frac{M_X^2}{{M^\text{SM} _Z}^2}\,.
\end{equation}
Here, $\theta_W$ denotes the weak mixing angle, $M_X$ is the bare mass of the new gauge boson and $M_Z^\text{SM}=\sqrt{g^2+g^{\prime 2}}\, v/2$ is the SM mass of the $Z$-boson.

\begin{table}[t]
\begin{center}
 \begin{tabular}{ c |  c   c   c    c   c   c } 
 \hline\\[-1.5ex]
  $f$ & $e$ & $\nu_e$  & $\mu, \nu_\mu$ & $\tau, \nu_\tau$ & $q_d$ & $q_u$ \\[1ex]
\hline\\[-1.5ex]
   $c_f$ &  $\epsilon\, e$  &  0 & $g_{\mu \tau}$ & $-g_{\mu \tau}$ & $\frac{1}{3}\epsilon\, e$ & $-\frac{2}{3}\epsilon\, e$ \\[1ex]
  \hline
\end{tabular}
\end{center}
\caption{Coupling coefficients $c_f$ for the interaction of the massive hidden photon $A'_\alpha$ to the fermionic vector current $\bar f \gamma^\alpha f$ of SM particles at leading order in the mixing parameter $\epsilon$.}\label{tab:cpls}
\end{table}

From \cref{eq:currentcouplings,eq:Kmatrix} we see that interactions of the massless photon state $A$ remain the same as in the SM. The massive hidden photon $A'$, however, couples both to the $L_\mu-L_\tau$ current $J^{\mu-\tau}_\alpha$ via gauge interactions and to the electromagnetic current $J^\text{EM}_\alpha$ via kinetic mixing. Hence, the interactions of the physical hidden photon $A'$  can generically be expressed as
\begin{equation} \label{eq:hpint}
\C{L}_{f A'} = - c_f\,\bar f \gamma^\alpha f\, A^{\prime}_{\alpha} \,,
\end{equation}
where the coupling coefficients $c_f$ for the different fermions $f$ are summarised in \cref{tab:cpls}.
 \par

In a simple \Umt extension of the SM, kinetic mixing can arise at tree level through the gauge invariant, renormalisable mixing term in~\cref{eq:lagr}. However, if the new gauge symmetry \Umt is embedded into a larger non-abelian symmetry structure in the UV (such as e.g.~$SU(N)_{L_\mu-L_\tau} \to U(1)_{L_\mu-L_\tau}$) a fundamental mixing term is forbidden by gauge invariance. Nevertheless, kinetic mixing will still be generated radiatively at the loop level and is hence unavoidable. For a general in-detail analysis of the origin of one-loop kinetic mixing we refer to the discussion in~\cref{sec:loopmix}. For the sake of clarity, we will restrict ourselves in the present work to the case where a tree-level mixing term is either forbidden by the larger symmetry structure of the UV theory, or is subdominant compared to the loop-induced mixing.

At low energies, $q^2\ll M_Z^2$, the kinetic mixing can be approximately considered to be only with the photon. In this regime,
the full one-loop result for the loop-induced  kinetic mixing parameter in the minimal \Umt setup reads, 
\begin{align}\label{eq:lmutaumix}
\epsilon_{\mu\tau}(q^2)=\frac{e\, g_{\mu \tau}}{2\pi^2}\int_0^1 dx\,x(1-x)\bigg[\log\bigg(\frac{m_\mu^2-x(1-x)q^2}{m_\tau^2-x(1-x)q^2}\bigg) \bigg]\,,
\end{align}
where  the fermions $\mu$ and $\tau$, charged under both \Umt and the SM $U(1)_\mathrm{EM}$, are running in the loop.
Here $q^2$ denotes the momentum transfer flowing through the loop. 
For very large momenta, $q^2\gg m_\tau^2$, the mixing is momentum suppressed, $\epsilon_{\mu\tau}\propto m_\mu^2/q^2 - m_\tau^2/q^2$. However, as long as the momentum transfer is below the muon mass $q^2 \ll m_\mu^2$, the mixing is approximately constant\footnote{It should be noted that the kinetic mixing parameter as defined in~\cref{eq:kinmixapprox} is negative, $\epsilon_{\mu\tau}<0$.},
\begin{equation}\label{eq:kinmixapprox}
    \epsilon_{\mu\tau}(q^2 \ll m_\mu^2)\approx \frac{e \, g_{\mu \tau}}{6\pi^2}\log\left(\frac{m_\mu}{m_\tau} \right)  \sim - \frac{g_{\mu\tau}}{70}  \,,
\end{equation}
which corresponds to the physical situation encountered in all the processes discussed in this work.

\subsection{Muon anomalous magnetic moment}
\label{sec:g-2}

\begin{figure}[t]
\begin{center}
\includegraphics[width=.4\textwidth]{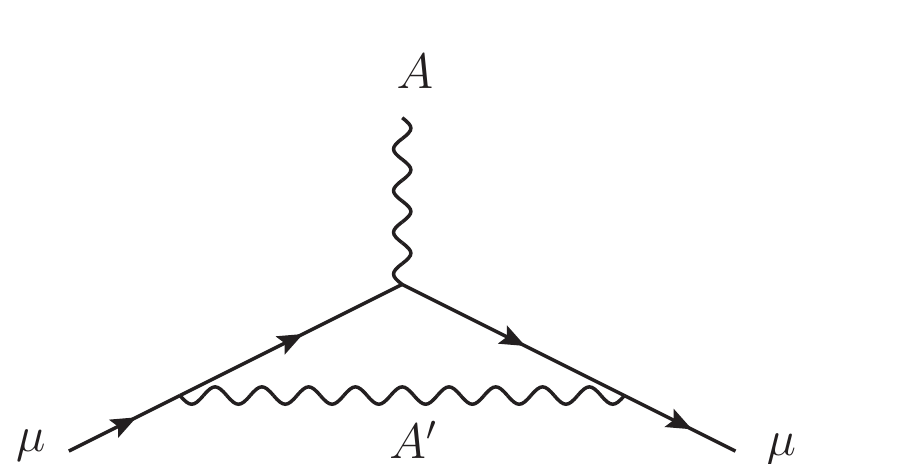}
\end{center}
\caption{\label{fig:g2} Contribution of the \Umt gauge boson to the anomalous magnetic moment of the muon.}
\end{figure}

Due to its gauge interactions with the second generation leptons the \Umt gauge boson $A'$ contributes to the anomalous magnetic moment of the muon, $a_\mu = (g-2)_\mu/2$, via the one-loop process displayed in~\cref{fig:g2}
\footnote{Due to non-zero kinetic mixing the \Umt gauge boson contributes also to the electron anomalous magnetic moment $(g-2)_e$ through the analogous process of~\cref{fig:g2} with the external muon legs replaced by electron legs. However, as we consider scenarios where kinetic mixing only arises at the one-loop level, the contribution to $(g-2)_e$ is effectively a three-loop process and hence very much suppressed. Nevertheless, the latest experimental determination of $(g-2)_e$ shows a mild deficit compared to the SM prediction~\cite{Parker191} and due to the positive sign the corresponding contribution via the \Umt boson slightly worsens this tension.
}. For any neutral gauge boson with vectorial couplings to muons (as in \Umt), the additional contribution to $a_\mu$ can be expressed in the compact form~\cite{Lynch:2001zs,Pospelov:2008zw},
\begin{equation}\label{eq:g2mu}
    \Delta a_\mu = Q_\mu^{\prime 2}\ \frac{\alpha'}{\pi} \int_0^1 du \ \frac{u^2(1-u)}{u^2+\frac{(1-u)}{x_\mu^2}}\,,
\end{equation}
where $\alpha'=g_{\mu \tau}^2/4\pi$, $x_\mu=m_\mu/M_A'$ and $Q'_\mu$ denotes the charge of the muon under the new gauge symmetry, i.e.~in the case of \Umt we have $Q'_\mu=1$.
\par
The experimentally determined value at BNL~\cite{Bennett:2002jb,Bennett:2004pv,Bennett:2006fi} and the latest theoretical evaluation of the SM prediction~\cite{Aoyama:2012wk,Aoyama:2019ryr,Czarnecki:2002nt,Gnendiger:2013pva,Davier:2017zfy,Keshavarzi:2018mgv,Colangelo:2018mtw,Hoferichter:2019gzf,Davier:2019can,Keshavarzi:2019abf,Kurz:2014wya,Melnikov:2003xd,Masjuan:2017tvw,Colangelo:2017fiz,Hoferichter:2018kwz,Gerardin:2019vio,Bijnens:2019ghy,Colangelo:2019uex,Blum:2019ugy,Colangelo:2014qya}\footnote{For details on the individual contributions entering the SM calculations we refer the reader to Ref.~\cite{Aoyama:2020ynm}.} for the anomalous magnetic moment of the muon read
\begin{align}
     a_\mu^\mathrm{exp} =  116\ 592\ 089(63) \times 10^{-11}\,, \\[.2cm]
     a_\mu^\mathrm{SM} =  116\ 591\ 810(43) \times 10^{-11}\,.
\end{align}
These values expose  the quoted $3.7\,\sigma$ deviation~\cite{Aoyama:2020ynm} of
\begin{equation}
    \Delta a_\mu = 279(76)\times 10^{-11}\,.
\end{equation}
Interpreting this deviation as a signal for new physics, the observed discrepancy can be explained by virtue of \cref{eq:g2mu} with the presence of a \Umt gauge boson for an appropriate choice of the gauge coupling strength $g_{\mu \tau}$.\par
It is worth noting that the additional contribution $\Delta a_\mu$ induced by vectorial couplings in~\cref{eq:g2mu} is strictly positive, regardless of the charge of the muon $Q'_\mu$ under the new symmetry. Hence, models with vectorial couplings can only explain an upward fluctuation in $(g-2)_\mu$. With new experimental results from E989 imminent it is worthwhile pointing out that in case of a possible observation of a downward fluctuation compared to the SM prediction, models with vectorial couplings to muons, like \Umt, would be in strong tension with the experimental findings and could be ruled out as an explanation due to the opposite sign of $\Delta a_\mu$.\par
In case the central value of the observed anomalous magnetic moment should move downward and would be in agreement with the SM prediction, $(g-2)_\mu$ would simply turn into a constraint and yield an upper bound on the gauge coupling constant $g_{\mu \tau}$ of the \Umt boson.

\subsection{Non-standard neutrino interactions}
\label{sec:nu-nsi}

The leptophilic \Umt gauge boson induces new  neutrino-matter interactions. If the mass of the new  mediator $A'$ is much heavier than the relevant energy scale of the scattering, $M_{A'}^2\gg q^2$, the mediator can be integrated out and the process be described by a four-fermion contact interaction. However, even in the regime where the mediator mass and the relevant scattering energy are comparable, $M_{A'}^2 \simeq q^2$, we can encapsulate the  hidden photon interaction in a quasi-effective (energy-dependent) contact term of the neutrinos $\nu_\alpha$ and fermions $f$. Making  use of the framework of neutrino non-standard interactions (NSI)~\cite{Davidson:2003ha,GonzalezGarcia:2004wg} this contact term can be written as\footnote{Note that here $\varepsilon^{fP}_{\alpha\alpha}$ denotes the strength of the effective NSI, not to be confused with the loop-induced kinetic mixing parameter $\epsilon_{\mu\tau}(q^2)$.},
\begin{equation}\label{eq:eff_op}
    \C{L}_\text{NSI} = -2\sqrt{2}\, G_F \sum_{\substack{f=u,d\\\alpha=e,\mu,\tau}} \varepsilon^{fP}_{\alpha\alpha} \ \left[\bar\nu_\alpha \gamma_\rho P_L \nu_\alpha\right] \, \left[\bar f \gamma^\rho P f \right]\,,
\end{equation}
where  $G_F$ denotes the Fermi constant and $P=\{P_L,P_R\}$. In order to match the gauge boson interaction onto that of the effective operator in~\cref{eq:eff_op}, we need to identify the amplitudes of the $\left[\bar\nu_\alpha \gamma_\rho P_L \nu_\alpha\right] \, \left[\bar f \gamma^\rho P f \right]$ interaction in the effective theory with the one in the full theory.
Making use of the expression for the four-momentum transfer, $q^2= - 2\, m_f\, E_R$, the matching condition yields an effective energy-dependent NSI coupling of
\begin{equation}\label{eq:nsi_cpls}
    \varepsilon^{fP}_{\alpha\alpha}(E_R) =  \frac{g_{\mu \tau}\, Q'_{\nu_\alpha}}{2\sqrt{2}\, G_F\,(2 m_f E_R+M_{A'}^2)} \, \times \begin{cases}{}
    - e\,\epsilon_{\mu\tau}\, Q_{f}^\text{EM}\,, & \text{for a kinetic mixing coupling to}\ f\,, \\[.5cm]
    g_{\mu \tau}\, Q'_{f} \,, & \text{for a gauge coupling to}\ f\,,
    \end{cases}
\end{equation}
where the minus sign in the kinetic mixing coupling is due to the relative sign difference  of the gauge and kinetic mixing interaction of the $A'$ in~\cref{eq:currentcouplings,eq:Kmatrix}. Furthermore, we  denote the corresponding neutrino charges under \Umt by $Q'_{\nu_\alpha} = \{0,1,-1\}$ for $\alpha=\{e, \mu , \tau\}$, respectively.\par
In the following we will use the NSI framework to study the effect of an additional \Umt gauge boson on a number of neutrino physics processes.
\bigskip

\begin{itemize}
\item{\textbf{Coherent elastic neutrino-nucleus scattering}}

The first neutrino interaction we are considering in this article that receives observable modifications in the presence of a \Umt gauge boson is  \coherent. 
In order to compute the $A'$ contribution to this process we can make use of the NSI formulation in~\cref{eq:eff_op}. This modifies  the low-energy formulation of the weak neutral currents to
\begin{align}\label{eq:nc}
    \C{L}_\text{NC} = - 2 \sqrt{2} \, G_F \sum_{P,f, \alpha} ( g^f_P+ \varepsilon^{fP}_{\alpha\alpha})\ [\bar \nu_\alpha \gamma_\rho P_L \nu_\alpha] \, [\bar f \gamma^\rho P f] \,,
\end{align}
where we denote the coupling of the SM $Z$-boson to the fermion $f$ as 
\begin{equation}\label{eq:Zcpls}
    g^f_P=T^3_f - \sin^2\theta_W\, Q^\text{EM}_f\,.
\end{equation}
Following the standard derivation of the cross section~\cite{Papoulias:2013gha,Papoulias:2015vxa}, we can write this parametrically  as 
\begin{align}\label{eq:sig_CEVNS_gen}
    \frac{d \sigma_{\nu_{\alpha\,N}}}{d E_R}  = \frac{G_F^2\, M_N}{\pi}\left(1-\frac{M_N\, E_R}{2 E_\nu^2}\right)\ \bigg\{ {G^\text{SM}_{\nu N}}^2 +2\, G^{\text{SM}}_{\nu N}\, G^{\text{NSI}}_{\nu_\alpha N} + {G^\text{NSI}_{\nu_\alpha N}}^2    \bigg\}\, F^2(E_R) \,,
\end{align}
where $E_\nu$ is the energy of the incoming neutrino, $E_R$ and $M_N$ are the recoil energy and mass of the scattered nucleus. For simplicity, we have assumed that the electromagnetic form factors $F_Z(q^2)$ and $F_N(q^2)$ describing the charge distribution within protons and neutrons are roughly given by the Helm nuclear form factor $F(E_R)$~\cite{Helm:1956zz,Lewin:1995rx}, which describes the nucleon distribution within a nucleus in terms of the magnitude of the three-momentum transfer $Q=\sqrt{2M_N E_R}$ as
\begin{equation}
  F(Q^2) = 3\frac{j_1(QR_0)}{QR_0}\exp(-Q^2s^2/2) ,
  \label{eq:form_factor}
\end{equation}
where $j_1$ denotes the spherical Bessel function of order one, and $R_0^2=c^2 + \frac73\pi^2a^2 - 5s^2$ with $c\simeq(1.23\, A^{1/3} - 0.6)$~fm, $a\simeq0.52$~fm and $s\simeq0.9$~fm.\footnote{We have checked that using the nucleon-specific form factors makes a negligible difference to our results.} 
\par
The relevant neutrino-nucleus scattering amplitudes corresponding to SM and NSI scattering, respectively, can be expressed as
\begin{align}\label{eq:gnuSM}
  G_{\nu N}^\text{SM}  &= \left(2 g^{u}_{V} + g^{d}_{V} \right) \ Z + \left(  g^{u}_{V} + 2 g^{d}_{V} \right)\ N 
  \,,\\[.3cm]
 G_{\nu_\alpha N}^\text{NSI}  &= \left(2 \varepsilon^{uV}_{\alpha\alpha}(E_R) + \varepsilon^{dV}_{\alpha\alpha}(E_R) \right)\, Z + \left( \varepsilon^{uV}_{\alpha\alpha}(E_R) + 2 \varepsilon^{dV}_{\alpha\alpha}(E_R) \right)\, N 
 \,, \label{eq:gnuNSI}
\end{align}
where we have made use of the vector couplings defined as
\begin{align}
g^f_V  &= g^f_L + g^f_R \,,\\[.3cm] 
    \varepsilon^{fV}_{\alpha\alpha} &= \varepsilon^{fL}_{\alpha\alpha}+\varepsilon^{fR}_{\alpha\alpha} \,.
\end{align}
Replacing our expressions for the $A'$-induced NSI couplings defined in~\cref{eq:nsi_cpls}, the full expression for the differential neutrino-nucleus scattering cross section can be written in the compact form,
\begin{multline}
    \frac{d \sigma_{\nu_{\alpha\,N}}}{d E_R}  = \frac{G_F^2\, M_N}{\pi}\left(1-\frac{M_N\, E_R}{2 E_\nu^2}\right)\\ 
    \times \,\bigg\{ \frac{Q_{\nu N}^2}{4} \, 
    +\, \frac{g_{\mu \tau}\,e\,\epsilon_{\mu\tau}\, Z\ Q'_{\nu_\alpha}\, Q_{\nu N} }{\sqrt{2}\, G_F\, (2 M_N E_R+ M_{A'}^2)} \, 
    + \, \frac{g_{\mu \tau}^2\,e^2\,\epsilon_{\mu\tau}^2\, Z^2\ Q^{\prime 2}_{\nu_\alpha} }{2\, G_F^2\, (2 M_N E_R+ M_{A'}^2)^2}    \bigg\}\, F^2(E_R) \,,
    \label{eq:sig_numu_mur}
\end{multline}
where we have defined the coherence factor for the effective neutrino-nucleus coupling via the SM $Z$-boson as
\begin{equation}
    Q_{\nu N} = N - (1 -4\,\sin^2\theta_W)\, Z \,.
\end{equation}
From~\cref{eq:sig_numu_mur} we can see that  the \Umt gauge boson couples  proportional to the total electric charge of the nucleus, $Z$. This is expected as in the case of \Umt the coupling to hadrons is generated from kinetic mixing and hence proportional to the photon coupling, i.e.~the electric charge\footnote{This is fundamentally different from models exhibiting gauge interactions with baryons, like e.g.~$U(1)_{B-L}$, where the boson couples proportional to the baryon number of the quarks (cf.~\cite{Boehm:2018sux}). There the coupling of the $A'$ to the nucleus is proportional to the total number of baryons bound in the nucleus, $A=Z+N$.}.
\newline\vspace{2ex}

%
%
\begin{figure}[t]
\begin{center}
\includegraphics[width=.32\textwidth]{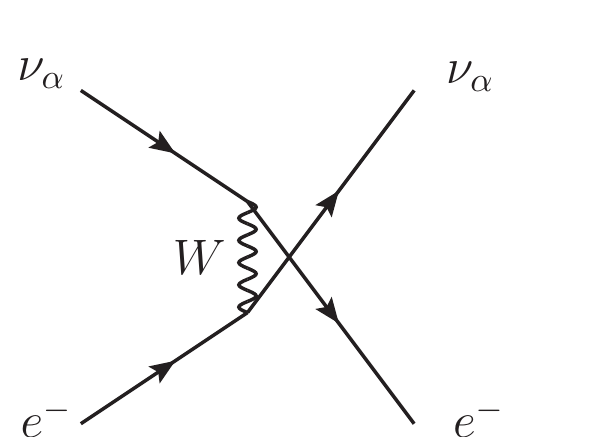}
\includegraphics[width=.32\textwidth]{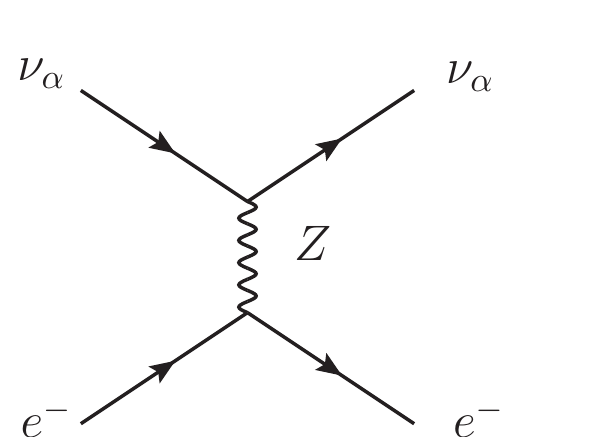}
\includegraphics[width=.32\textwidth]{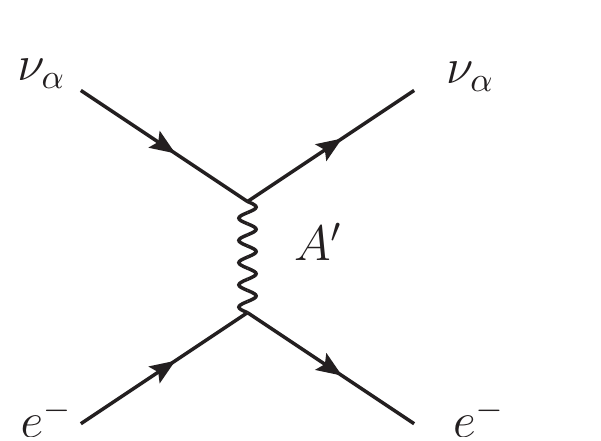}
\end{center}
\caption{\label{fig:nue} Tree-level contributions to the scattering $\nu_\alpha e \to \nu_\alpha e$. The first diagram only contributes to the process for $\alpha=e$, whereas the last one contributes only for the flavours $\alpha=\mu,\tau$.}
\end{figure}
%
%

\item{\textbf{Elastic neutrino-electron scattering}}

The second process we consider to probe neutrino interactions of a new \Umt 
is elastic neutrino-electron scattering, $\nu_\alpha e \to \nu_\alpha e$. In the SM, neutrino-electron scattering proceeds via the weak interactions.
For the neutrino flavours $\alpha=\mu, \tau$ this process  is solely mediated by the neutral current interaction in~\cref{eq:nc} (corresponding to the diagrams in the centre and right panel of~\cref{fig:nue}). However, for $\alpha=e$ this process is also  mediated by a charged current interaction (corresponding to the diagram in the left panel of~\cref{fig:nue}). The corresponding low-energy charged current  contact term reads
\begin{align}\label{eq:cc}
    \C{L}_\text{CC} = - 2 \sqrt{2} \, G_F\ [\bar \nu_\alpha \gamma_\rho P_L \ell] \, [\bar \ell \gamma^\rho P_L  \nu_\alpha] \,.
\end{align}
The general form of the neutrino-electron scattering cross section induced by these low-energy neutral and charged current interactions  can then be expressed as~\cite{Radel:1993sw,Marciano:2003eq,Bolanos:2008km,Formaggio:2013kya,Lindner:2018kjo,Dutta:2019oaj},
\begin{equation}\label{eq:sig_el}
  \frac{d \sigma_{\nu_{\alpha\,e}}}{d E_R}  = \frac{2 G_F^2\, m_e}{\pi} \ \left[ (g_1^\alpha)^2+(g_2^\alpha)^2\left(1-\frac{E_R}{E_\nu}\right)^2 - g_1^\alpha\, g_2^\alpha\, \frac{m_e  E_R}{E_\nu^2} \right]\,,    
\end{equation}
where once more $E_\nu$ denotes the initial energy of the incoming neutrino and $E_R$ the  recoil energy of the scattered electron.
In the SM the relevant weak interaction couplings are given by 
\begin{align}\label{eq:cpls_esm}
g_1^\alpha=
\begin{cases} 1 + g_L^e\,, & \text{for}\ \alpha = e\,, \\
g_L^e\,, & \text{for}\ \alpha = \mu,\tau\,,
\end{cases}  &&
g_2^\alpha = g_R^e\,, \quad \text{for}\ \alpha = e,\mu,\tau \,,
\end{align}
where $g_L^e$ and $g_R^e$ refer to the couplings of the $Z$-boson to electrons as defined in~\cref{eq:Zcpls}.
It is now straightforward to extend~\cref{eq:sig_el} to incorporate  neutrino-electron scattering mediated by the new \Umt. By virtue of the NSI formulation of the $A'$ interaction, we can simply make the replacement 
\begin{align}
    g_1^\alpha \to g_1^\alpha + \varepsilon^{eL}_{\alpha\alpha} \,, &&
    g_2^\alpha \to g_2^\alpha + \varepsilon^{eR}_{\alpha\alpha} \,,
\end{align}
where $\varepsilon^{eP}_{\alpha\alpha}$ are the $A'$-induced NSI couplings to electrons defined in~\cref{eq:nsi_cpls}. As the first generation of leptons is uncharged under \Umt the cross section for $\nu_e \, e \to \nu_e \, e$ scattering will to lowest order remain unchanged. However, for $\alpha=\mu,\tau$ the scattering cross section for the process $\nu_\alpha\, e \to \nu_\alpha\, e$ is modified and explicitly reads
\begin{align}\label{eq:sig_el_nsi}
    \frac{d \sigma_{\nu_{\alpha\,e}}}{d E_R}  =   &\frac{2\, G_F^2\,m_e}{\pi} \ \Bigg\{ \,  \left[ {g^e_L}^2+{g^e_R}^2\left(1-\frac{E_R}{E_\nu}\right)^2 - g^e_L\,g^e_R \frac{m_e \, E_R}{E_\nu^2} \right]  \notag \\
    &+ \frac{g_{\mu \tau}\, e\, \epsilon_{\mu\tau} \,  Q'_{\nu_\alpha}}{\sqrt{2}\, G_F (2E_R\, m_e + M_{A'}^2)} \left[ (g^e_L+g^e_R)\left(1-\frac{m_e\, E_R}{2E^2_\nu}\right) - g^e_R \frac{ E_R}{E_\nu}\left(2- \frac{E_R}{E_\nu}\right) \right]  \notag \\
    &+  \frac{g_{\mu \tau}^2 \, e^2\, \epsilon^2_{\mu\tau} \,  Q^{'2}_{\nu_\alpha}}{\,4\, G_F^2 (2E_R\, m_e + M_{A'}^2)^2} \left[  1 - \frac{ E_R}{E_\nu} \left(1 - \frac{E_R-m_e}{  2\, E_\nu}\right) \right]\, \Bigg\} \,.
\end{align}
The first term in the curly braces corresponds to the pure SM, the second to the interference and the last one to the pure BSM contribution. 
The sign of the interference term critically depends on the charge  $Q'_{\nu_\alpha}$ of the scattered neutrino $\nu_\alpha$.
\end{itemize}
\bigskip

Apart from the two types of elastic neutrino scattering processes just discussed, quite universal limits on NSI couplings are typically derived from the induced neutrino-matter potentials, which can be probed e.g.~in neutrino oscillation experiments~\cite{Farzan:2017xzy,Esteban:2018ppq}. These constraints are for example quite stringent on hidden photon models featuring gauge couplings to baryons~\cite{Khan:2019jvr,Flores:2020lji,Kling:2020iar}. However, in the case of \Umt the hidden photon couplings  to matter (i.e.~$p, n$ and $e$) arise only via kinetic mixing and are therefore proportional to the electric charge. Hence, as the matter within the Earth or experimental apparatuses is electrically neutral, the net effect of NSI couplings proportional to the electric charge is exactly zero as they  cancel out~\cite{Heeck:2018nzc},
\begin{equation}\label{eq:nsi_cancellation}
    \epsilon^p_{\alpha\beta} + \epsilon^e_{\alpha\beta} = 0\,, \qquad  \epsilon^n_{\alpha\beta} =0 \,.
\end{equation}
Therefore, the typically strong limits from NSI matter potentials do not apply in the case of a \Umt gauge boson. \par

Nevertheless, NSIs generated in the \Umt model still have the potential to affect global fits of neutrino mixing parameters, as they directly change the neutrino-electron elastic scattering cross section. This is important, as some of the main results used to constrain the mixing parameters $\theta_{12}$ and $\Delta m_{12}$ are the measurements of the scattering of $\mathrm{^8B}$ neutrinos with electrons by the Super-Kamiokande, SNO+ and Borexino experiments \cite{Agostini:2017cav,Aharmim:2011vm,Abe:2016nxk}. Any new physics which affects the electron-neutrino scattering rate will also affect this result \cite{Gonzalez-Garcia:2013usa}, leading to an incorrect determination of these neutrino mixing parameters. This kind of degeneracy has been explored in the context of quark NSIs \cite{Esteban:2018ppq}, but it has the potential to be much more significant with electron NSIs.

Two potential ways to resolve this degeneracy are apparent. Firstly, the effects of changing the neutrino mixing parameters and of introducing non-zero NSIs are likely to be different for measurements taken in different energy windows. All of the $\mathrm{^8B}$ neutrinos measured at Super-Kamiokande had energies above 5~MeV. This means that they had passed through the MSW resonance and so the probabilities of finding them in each neutrino flavour eigenstate will be different from those for lower energy solar neutrinos, while an energy dependence is also found in the neutrino-electron cross section. Secondly, for the case of a \Umt model the NSIs generated with electron-neutrinos are significantly suppressed relative to muon and tau neutrinos. The other major result used in calculating the quoted neutrino mixing parameters is the one from KamLAND, which detects reactor neutrinos by inverse-beta decay \cite{Gando:2013nba}. This means it is only able to detect electron anti-neutrinos, and is therefore unaffected by the NSIs $\epsilon^e_{\mu \mu}$ and $\epsilon^e_{\tau \tau}$.

In this context, it is most interesting to note that a tension currently exists between the mixing parameters computed using only KamLAND data and those using only solar neutrino data, especially in the value of $\Delta m_{12}$ \cite{Esteban:2018azc}. 
It seems likely that the only consistent way to obtain a simultaneously correct treatment of both neutrino mixing and  NSIs with quarks and electrons, suitable to determine whether NSIs can resolve the tension in measurements of $\Delta m_{12}$, would be to include all the mixing parameters and NSI couplings in a single global fit analysis. 
Once such a holistic global fit including all NSI parameters will be available, it will allow for an improved analysis of the limits.
In the absence of such a fit and for the remainder of this work we take our neutrino mixing parameters from the global fit in Ref.~\cite{Esteban:2018azc}.

\section{Solar neutrino probes of the gauged \Umt}
\label{sec:constraints}

As we have seen in the previous section, the \Umt hidden photon introduces new neutrino interactions that have an effect on neutrino electron-scattering and coherent neutrino-nucleus scattering. In this section we will study how these effects can be probed. We start by considering neutrino-electron scattering in dedicated neutrino experiments and reevaluate constraints from the Borexino detector.
Then, we consider coherent neutrino-nucleus scattering in spallation facilities and derive new limits from the recent CENNS-10 LAr results. Finally, we address dark matter experiments sensitive to both electron and coherent-nucleus scattering. In this context, we first derive limits from the recent XENON1T electron recoil data before we study the prospects for a number of upcoming and future direct detection experiments.

These three types of experimental techniques cover the parameter space of the gauged \Umt model in complementary ways. We will now explore how effectively they can probe the area of the parameter space that is consistent with a solution to the muon anomalous magnetic moment.

In the case of Borexino and the various dark matter experiments we will consider, the most relevant source of neutrinos is the Sun. Solar neutrinos can be divided into several individual flux spectra, each associated with a different stage in the fusion process. The most important fluxes for us are those generated during the proton-proton ($pp$) chain: the $pp$, $\mathrm{^7Be}$, $pep$, $\mathrm{^8B}$ and $hep$ neutrinos. Each of these spectra has a different total flux, which must be properly modelled for us to be able to place reliable constraints. The values we use for the individual neutrino fluxes are computed in Ref. \cite{Vinyoles:2016djt} within the context of the Standard Solar Model (SSM). Although these various neutrino fluxes are all produced in the electron neutrino eigenstate, they will undergo flavour oscillations as they traverse the Sun and travel to a detector on Earth. These neutrino oscillations are discussed in \cref{sec:oscillations}.

A significant source of uncertainty in the SSM prediction is the as-yet unsolved solar metallicity problem \cite{Cerdeno:2017xxl,Bahcall:2004yr,Bahcall:2005va,Bi:2011sy,Basu:2007fp}. The rates of various fusion processes depend on the metal content\footnote{Here we use the word metal in the astrophysical sense, meaning any element heavier than helium.} in the Sun. Two possible scenarios exist: the so-called high metallicity (HZ) GS98, and low metallicity (LZ) AGSS09 models \cite{Vinyoles:2016djt}. The solar metallicity has a significant effect on the rate of the Carbon-Nitrogen-Oxygen (CNO) fusion cycle: the associated CNO neutrino fluxes are around 50\% larger in the HZ scenario \cite{Vinyoles:2016djt}, and a direct measurement of these as-yet unobserved neutrinos could help to finally resolve the solar metallicity problem. However, it also affects the various $pp$-chain fluxes by up to 10\%. The reach of future experiments to explore new physics will therefore differ in these two scenarios. Constraints obtained from current results such as Borexino can differ even more: what appears like a small upwards fluctuation above the SM result under the assumption of a HZ Sun can be a much larger enhancement if the Sun has low metallicity, which could allow for a larger contribution to the interaction rate from new physics.\par
In order to maintain full generality when deriving limits from solar neutrino scattering we will hence explicitly perform our calculations for both the scenario of a high and low meallicity Sun.

\subsection{Borexino}
\label{sec:constraints_borexino}

With its large fiducial volume and low background rates, the Borexino experiment has been responsible for many of the most precise direct measurements to-date of the various solar neutrino fluxes through their scattering with electrons in a scintillating liquid target. With analyses covering energies from 0.19 to 20~MeV, it has measured or placed constraints on the fluxes of $pp$, $\mathrm{^7Be}$, $pep$, $\mathrm{^8B}$, $hep$, and CNO neutrinos \cite{Agostini:2017ixy,Agostini:2017cav}, leading to constraints on models of new physics which affect the neutrino-electron scattering rate.

Among the various sources of solar neutrinos, Borexino has measured the $\mathrm{^7Be}$ rate to the highest level of precision. This, combined with the fact that the greatest change to the cross section from a new light mediator is at low energy, makes $\mathrm{^7Be}$ neutrinos the best candidate for placing constraints on the gauged \Umt model\footnote{The $\mathrm{^8B}$ spectrum has also been measured to high precision. However, as discussed in Section \ref{sec:nu-nsi}, these measurements are one of the key ingredients used to calculate the neutrino mixing parameters that we use as inputs, and so should not be used again to directly constrain our model.}.

In Ref. \cite{Bellini:2011rx}, the Borexino collaboration measured the $\mathrm{^7Be}$ flux with a precision of 5\%, finding it to be in agreement with the SM and SSM. This result has been used in an earlier work to place constraints on a model of gauged $U(1)_{B-L}$, excluding all regions of parameter space where the enhancement of the flux over the SM prediction exceeds 8\% (the corresponding 90\% CL given a $1\sigma$ uncertainty of 5\%) \cite{Harnik:2012ni}.  This result has itself been used to place a constraint on a \Umt model by remapping the associated couplings \cite{Kaneta:2016uyt}. We have improved on this method in several ways. Firstly, by calculating constraints directly for a \Umt model we are able to take into account the full scattering cross section from~\cref{eq:sig_el_nsi} including the interference term. Secondly, the original $B-L$ result did not take into account the uncertainty on the theoretical prediction from the SM and SSM, including the uncertainty stemming from the solar metallicity problem. Finally, results from Phase II of Borexino have since measured the flux with even higher precision, of around 2.7\% \cite{Agostini:2017ixy}. We calculate constraints on a \Umt model from both Phase I and II Borexino results in both metallicity scenarios using a chi-squared test. In~\cref{sec:b-l} the same process is used to derive updated constraints on the $U(1)_{B-L}$ model for which these constraints have been first derived.

Under the assumption of high solar metallicity, and based on the work of \cite{Vinyoles:2016djt}, the Borexino collaboration computed a theoretical uncertainty on the $\mathrm{^7Be}$ rate of 5.8\%. Combining this with the uncertainty on the experimental measurements, we derive constraints at the 90\% confidence level using a chi squared test. While the two Borexino results are consistent with each other and with the SM (under the assumption of a high metallicity SSM), the Borexino Phase II measured a higher value for the $\mathrm{^7Be}$ flux, and so gives weaker constraints on our model. As these two results can be consistently explained as a downwards and upwards fluctuation, respectively, we will display the constraints derived from each analysis, with the assumption that a more sophisticated analysis would lie somewhere between the two.

The predicted $\mathrm{^7Be}$ neutrino flux is $\sim10$\% lower under the assumption of a low metallicity Sun than it is with a high metallicity Sun. A larger contribution to neutrino-electron scattering from new physics is therefore allowed by the measurements from Borexino.

A more complete method of deriving constraints on a \Umt model from all the solar neutrino measurements from Borexino would be to combine them all in a single chi-squared analysis. However, as the best fit values for the count rates of $pp$, $\mathrm{^7Be}$, and $pep$ were computed simultaneously in a single multivariate fit \cite{Agostini:2017ixy}, a careful handling of the various uncertainties would be required, and we leave this to future works.

\subsection{Coherent neutrino-nucleus scattering at COHERENT}
\label{sec:constraints_coherent}

Having considered dedicated searches for neutrino-electron scattering, we turn our attention to experiments which are specifically designed to measure coherent neutrino-nucleus scattering. 
The COHERENT collaboration observed this elusive process for the first time in CsI \cite{Akimov:2017ade} and, more recently, measured it again using a liquid argon target (in the CENNS-10 LAr detector \cite{Akimov:2020pdx}) at the Oak Ridge National Laboratory Spallation Neutron Source. The results are compatible with the SM prediction and therefore can be interpreted as constraints on any new physics contribution, which are especially relevant for models with light mediators~\cite{Abdullah:2018ykz,Banerjee:2018eaf,Papoulias:2019txv,Khan:2019cvi, Miranda:2020tif}. For completeness, we have incorporated the limits from the CsI run, derived in Ref.~\cite{Abdullah:2018ykz}. To derive the constraints from the new LAr data, we proceed as follows.

The number of expected \coherent events can be expressed as
\begin{equation}
\begin{aligned}
N_{{\rm CE}\nu{\rm NS}} = \sum_{\nu_\alpha}  N_{\mathrm{targ}} \int_{E_{\mathrm{th}}}^{E_{R}^{\mathrm{max}}} \int_{E_{\nu}^{\mathrm{min}}}^{E_{\nu}^{\mathrm{max}}} \frac{\mathrm{dN_{\nu _\alpha}} }{\mathrm{d}E_\nu }
\mathcal{ \epsilon } ( E_R)  \frac{d \sigma_{\nu_{\alpha\,N}}}{d E_R}
dE_\nu dE_R \, ,
\end{aligned}
\label{eq:coherent_events}
\end{equation}
where ${\mathrm{dN_{\nu _\alpha}} }/{\mathrm{d}E_\nu }$ refers to the spectrum of the incoming neutrino $\nu_\alpha$ produced at SNS, the minimum and maximum neutrino energies are  $E_{\nu}^{\mathrm{min}}\approx\sqrt{M_N E_R/2}$ and $E_{\nu}^{\mathrm{max}}=m_\mu/2$, and
the expression for the differential cross-section is that of~\cref{eq:sig_numu_mur}. The number of target nuclei is $N_{\mathrm{targ}}={M_{tot}}/{M_N}$, with $M_{tot}=24$~kg (the exposure of CENNS-10 LAr) and $M_N$ the atomic mass of $^{40}$Ar, which we consider (for simplicity) to have 100\% isotopic abundance. For the energy-dependent efficiency, $\epsilon(E_R)$, we take the one  given in analysis A of Ref.~\cite{Akimov:2020pdx}.

The spectrum of stopped-pion neutrinos produced at SNS  
 contains a delayed neutrino flux, composed of muon antineutrinos and electron neutrinos, both with a maximum neutrino energy of $E_\nu^{\rm max}=m_\mu/2$ (see e.g., Refs.~\cite{Akimov:2018vzs,Akimov:2017ade}) and fluxes given by
\begin{eqnarray}
\frac{\mathrm{dN_{\overline{\nu} _\mu }} }{\mathrm{d}E_\nu }&=& \eta \frac{64E_\nu^{2}}{m_{\mu }^{3}}\left ( \frac{3}{4}-\frac{E_\nu}{m_\mu } \right ) \, ,  \nonumber
\\
\label{eq:flux_numu}
\frac{\mathrm{dN_{\nu _e }} }{\mathrm{d}E_\nu }&=& \eta \frac{192E_\nu^{2}}{m_{\mu }^{3}}\left ( \frac{1}{2}-\frac{E_\nu}{m_{\mu }} \right ) \, , 
  \nonumber
\label{eq:flux_nue}
\end{eqnarray}
as well as a monoenergetic muon neutrino flux, with $E_\nu=29.8$~MeV,
\begin{equation}
\frac{\mathrm{dN_{\nu _\mu }} }{\mathrm{d}E_\nu }=\eta\delta\left ( E_\nu-\frac{m_{\pi }^{2}-m_{\mu }^{2}}{2m_{\pi }} \right ) \, .
  \nonumber
\label{eq:flux_prompt}
\end{equation}
In these expressions, the normalisation factor $\eta = rN_{\text{POT}}/4\pi L^{2}$ takes into account the number of neutrinos of each type produced from each proton on target (POT) (the 6.12~GWh exposure corresponds to  $N_{\text{POT}}= 1.37\times10^{23}$). The number of neutrinos produced per POT is given by $r=0.08$, while $L=27.5$~m is the CENNS-10 baseline. In order to interpret the experimental data we need to translate from nuclear recoil energies to the equivalent energies of electron recoils. The quenching factor allowing for this conversion  from nuclear recoil energies to electron-equivalent energies is parametrised as $Q_F=0.246+7.8\times 10^{-4}\ E_R$, such that $E[\si{\kilo\electronvolt}\ee{}]=Q_F\, E_R$ \cite{Akimov:2020pdx}.

In order to derive the corresponding limit from the CENNS-10 LAr run on the parameter space of a \Umt gauge boson we perform a chi squared test.
For each point in the $(M_{A^\prime}, g_{\mu \tau})$ parameter space, the $\chi^2$ is then computed as follows \cite{Papoulias:2019lfi,Miranda:2020tif}
\begin{equation}
\begin{aligned}
\chi^2 (M_{A^\prime},g_{\mu \tau}) =  \underset{\alpha}{\mathrm{min}} \Bigg [ \left(\frac{N_{\mathrm{meas}} - N_{{\rm CE}\nu{\rm NS}}(M_{A^\prime},g_{\mu \tau})  [1+\alpha]}{\sqrt{N_{\mathrm{meas}} + N_{\rm bgd}}}\right)^2 
  + \left(\frac{\alpha}{\sigma_\alpha} \right)^2  \Bigg ] \, ,
\end{aligned}
  \nonumber
\label{eq:chi}
\end{equation}
where the number of observed events in the detector is $N_{\mathrm{meas}}=159 \pm 43$ (stat.)$ \pm 14$ (syst.), the number of background events is $N_{\rm bgd}=563$, and $\sigma_\alpha=0.085$ accounts for a scaling systematic.

\subsection{Direct dark matter detection experiments}
\label{sec:constraints_dm}

Dark matter direct detection experiments are, in principle, also sensitive to MeV-scale neutrino interactions through both their scattering off electrons and their coherent scattering off nuclei~\cite{Cerdeno:2016sfi,Essig:2018tss,Gonzalez-Garcia:2018dep}. Although their performance is still limited by the somewhat smaller target size than dedicated neutrino experiments, the next generation of detectors with extremely low energy threshold (such as SuperCDMS) or increased fiducial volume (liquid noble gas detectors such as LZ or DarkSide) will provide a complementary test of new physics in the neutrino sector.

The original focus of most of the detectors was to search for nuclear recoils (NR) for which the experimental design and data analysis generally guarantees a very small (ideally zero) background thanks to a very careful rejection of the large electron recoil background. However, in the last decade a great effort has been made to exploit electron recoils (ER) for dark matter (and other new physics) searches. Electron recoils allow to probe lighter particles but, in general, at the expense of introducing a larger background. In our analysis we consider both types of searches, as they explore the parameter space of new physics in the neutrino sector in a synergic manner \cite{Cerdeno:2016sfi}. In fact, when we give up on discrimination between both types of events, one can do a combined (NR+ER) analysis in which the signals (and backgrounds) from both types of recoils are added. This can be advantageous to improve the threshold for nuclear recoils, so we also take it into account.

In full analogy to~\cref{eq:coherent_events} we write the expected number of solar neutrino-electron or -nucleus scattering events at direct detection experiments as
\begin{equation}
    \label{eq:n_dd}
    N = \varepsilon\, n_T \int_{E_\mathrm{th}}^{E_{\mathrm{max}}} \sum_{\nu_\alpha}\int_{E_\nu^\mathrm{min}} \frac{d \phi_{\nu_e}}{d E_\nu}\ P(\nu_e \rightarrow \nu_\alpha)\  \frac{d \sigma_{\nu_{\alpha\,T}}}{d E_R} \ d E_\nu d E_R\, ,
\end{equation}
where now $E_{\nu}^{\mathrm{min}} = (E_R + \sqrt{E_R^2 + 2 m_T E_R} )/2$  is the minimum neutrino energy to produce a nuclear recoil or electronic recoil of energy $E_R$, $m_T$ is the mass of the target, $n_T$ is the total number of targets (electrons or nuclei) per unit mass\footnote{For the electron case, we have taken a step-function approach to model the number of electrons free to scatter. Once the energy of the incoming neutrino is above the binding energy of an electron's orbital, that electron is deemed available to interact and does so via the free-electron approximation. This simple approach is sufficient for our analysis.}, and $d \sigma_{\nu_\alpha\, T}/d E_R$ is the differential cross section of the elastic scattering of neutrinos off electrons (\cref{eq:sig_el_nsi}) or nuclei (\cref{eq:sig_numu_mur}). We sum across all neutrino flavours, $\alpha=e,\,\mu,\,\tau$, and the incoming (solar) neutrino flux, $d \phi_{\nu_e}/d E_\nu$, is multiplied by the transition probability between the source and the detector $P(\nu_e \rightarrow \nu_\alpha)$.
For simplicity, we will consider the experimental exposure, $\varepsilon$, to be energy-independent. The threshold energy, $E_\mathrm{th}$, and maximum energy $E_{\mathrm{max}}$ are specific to the experiment.

So far, none of the results published by direct detection experiments probe new areas of the parameter space of the \Umt model. For this reason, in our analysis we have decided to focus on the potential of some representative experiments which are currently being deployed (SuperCDMS and LZ) and on third generation proposals (DarkSide-20k and DARWIN).
However, the most sensitive search for electron recoils has been recently released by the XENON1T collaboration \cite{Aprile:2020tmw}, which seems to hint at a possible excess at low energies. Due to its relevance and timeliness, we have decided to include this data in our analysis as well.

\subsubsection{XENON1T}
\label{sec:xenon1t}

The XENON1T collaboration \cite{Aprile:2017aty} operated a liquid xenon time projection chamber (with an active target of 2 tonnes) from 2016 to 2018 to look for exotic events at the Laboratori Nazionali del Gran Sasso (LNGS). Particle interactions within the detector produce a prompt scintillation (S1) signal, accompanied by a delayed electroluminescence (S2) signal, both detected by an array of photomultipliers. The ratio of both signals (S2/S1) can be used to discriminate electron from nuclear recoils, thereby guaranteeing a low background for the latter in the canonical analysis mode (which regards electron recoils as background).

In a recent analysis~\cite{Aprile:2020tmw} the data for electron recoils in XENON1T Science Run 1 has been analysed to look for potential contributions from new physics. The greatest challenge of this study is the existence of an irreducible background dominated at low energies by the $\beta$ decays of $^{214}$Pb and with additional contributions from other sources of contamination as well as the xenon target itself. The different contributions to the background are determined using a global fit in the region $1-210$~keV to obtain a best fit model, denoted as $B_0$.

\begin{figure}[t]
\begin{center}
\includegraphics[width=0.7\textwidth]{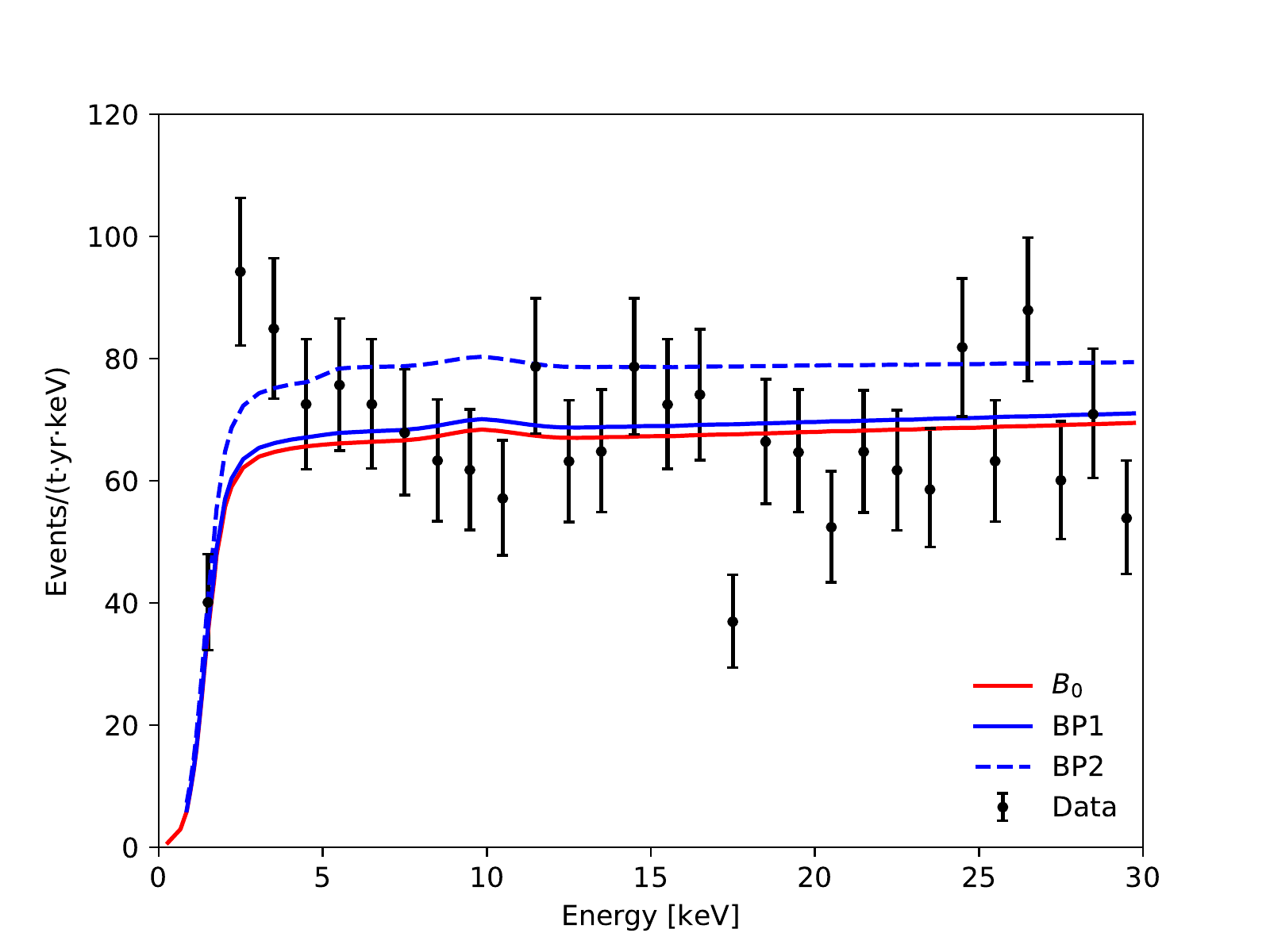}
\end{center}
\caption{\label{fig:xenon_spectrum} The data points represent the observed electron recoil spectrum at XENON1T. The red line corresponds to the fitted background model. The blue solid and dashed lines show two example spectra obtained with a \Umt gauge boson with mass $M_{A'}=15$~MeV and coupling $g_{\mu \tau}=5.6\times 10^{-4}$ (BP1) and $g_{\mu \tau}=1.7\times 10^{-3}$ (BP2), respectively.}
\end{figure}

The results obtained by the XENON1T collaboration are shown in~\cref{fig:xenon_spectrum}, together with the best fit model for the background (in red). An excess is observed in the low energy region (below 7~keV and more prominent in the first two bins) which can be understood as a hint for new physics (for example, solar axions, a neutrino magnetic moment, or light bosonic dark matter). However, as the authors of~\cite{Aprile:2020tmw} point out, a plausible explanation is also a previously unconsidered source of background, caused by traces of tritium (which undergoes $\beta$ decay) in the detector.

The enhancement of the spectrum over background attributed to an additional \Umt boson in our region of interest of the parameter space would be mostly flat over the energy range explored by XENON1T. In~\cref{fig:xenon_spectrum} we choose two benchmark points to show the enhancement to the recoil spectrum produced by a \Umt model over the top of the best-fit background spectrum from Ref. \cite{Aprile:2020tmw}, $B_0$. For the first benchmark point, BP1, we take $M_{A'}=15$~MeV and $g_{\mu \tau}=5.6\times 10^{-4}$, which lies in the region which can simultaneously resolve both the $H_0$ and $(g-2)_\mu$ anomalies, and is not excluded by any previous bounds. For the second, BP2, we take the same mass but increase the coupling to $g_{\mu \tau}=1.7\times 10^{-3}$, which is the point at which our model disagrees with the data at the 90\% confidence level (according to the calculation described below).
This means that a \Umt boson cannot explain the observed fluctuation in the XENON1T data, which is peaking in the region between $2-3$~keV. A \Umt or other light vector mediator could produce a feature resembling the XENON1T data, but this would require a much smaller mass, below $100$~keV \cite{Cerdeno:2016sfi}, a region of the parameter space which is already well-constrained in our model~\cite{Escudero:2019gzq,Poddar:2019wvu,Dror:2020fbh}.
\par
Therefore, we use the data from XENON1T to derive an approximate upper limit on the parameter space of the \Umt boson. Since our spectrum is relatively flat and featureless, we perform a simple unbinned delta chi-squared ($\Delta \chi^2$) analysis. Summing the data points, we compute the total number of events observed in the energy range $2-30$~keV, $N_{\mathrm{tot}}$. We then compare this value with the number of events predicted in the $B_0$ background model, and in the $B_0 + $\Umt scenario. We then place an upper limit at the 90\% confidence level using a $\Delta \chi^2$ test. 

For the purpose of performing the chi squared test we use the number of background events as determined in the background-only fit  by the Xenon collaboration. 
However, it should be noted that there is a large systematic uncertainty on the number of expected background events coming from $\beta$ decays of  $^{214}$Pb, quoted as $N^\mathrm{exp}_{\mathrm{Pb}}=(3450, 8530)$. This is particularly interesting as this constitutes by far the most dominant background in the low energy part of the spectrum, where e.g.~the number of background events from SM solar neutrinos $N^\mathrm{exp}_{S\nu}=220.7 \pm 6.6$ is more than an order of magnitude smaller. Due to the poor knowledge of the lead background it is fitted to data in the background-only model. Taking this as a fixed background for the purpose of our chi squared test with additional solar neutrino scattering due to a \Umt boson is overly simplistic, as any additional events from hidden photon-induced solar neutrino scattering could be easily compensated for by a fluctuation in the lead background. However, the approximation we take will lead to an overestimate of the severity of the constraint, and since the limit we derive does not constrain any new regions of the parameter space we can be confident that any limits derived from this XENON1T result will not be competitive with existing constraints. We also perform an equivalent analysis of a $U(1)_{B-L}$ model, and as in the \Umt case we find that limits from this result are not competitive (see \cref{fig:bl_lims}). \par
In a more careful derivation of the limit from the XENON1T result one should perform a full profile likelihood analysis, allowing the backgrounds to vary within the expected range as nuisance parameters. However, such a dedicated statistical analysis of the limits is beyond the scope of this sensitivity study.

\subsubsection{Forthcoming and future detectors}
\label{sec:future_dm}

Let us now turn our attention to the future sensitivities of dark matter experiments.
We consider simplified setups based on the characteristics of some prominent current detectors (to which we will refer as second generation, or G2) and  extend the analysis to some next generation (G3) proposals.
The experimental configurations are shown in Table\,\ref{table:exps}, where we specify the detector target, exposure, and energy ranges used in each analysis. Due to the different energy-loss mechanisms associated with each type of recoil event, the signals produced by electron and nuclear recoils must be interpreted differently when translating them into recoil energies. We differentiate between nuclear-recoil energies (denoted by \si{\kilo\electronvolt}\nr{}) and electron-equivalent energies (denoted by \si{\kilo\electronvolt}\ee{}).

In general, far more background events are expected from electron recoils (ER) than nuclear recoils (NR). A common way of reducing backgrounds in searches for nuclear recoils is through the combination of multiple detection channels, with nuclear and electron recoils having different profiles when viewed through both simultaneously. However, this typically limits the minimum energy threshold of the analysis. We consider three classes of analyses, denoted in \cref{table:exps} as NR, ER, and NR + ER. In the first two, our energy ranges are chosen to allow maximum discrimination between NR and ER events, and the signals of each recoil type are studied separately. In the latter, the two signals are combined, sacrificing a low-background nuclear recoil analysis to allow us to probe lower energy recoils. In this last type of analysis, we must first convert all recoil energies to the nuclear-recoil equivalent before combining our signals.

The nuclear recoil spectra are integrated up to the maximum kinematically allowed energies for the NR analyses, which follow from a collision with the highest energy solar neutrinos available (here the tail of the $hep$ neutrinos). Finally, we have not included energy resolution effects.

The G2-Ge setup is based on the planned configuration of SuperCDMS at SNOLAB \cite{Agnese:2016cpb}. The SuperCDMS experiment includes two types of detectors: iZIPs, which allow for an excellent discrimination between electron and nuclear recoils, and HV \cite{Agnese:2017jvy}, which operate at high voltage to exploit phonon  amplification and achieve a much lower energy threshold at the expense of not discriminating between electron and nuclear events
\footnote{In our analysis we have ignored the Silicon targets in SuperCDMS SNOLAB, as the larger exposure in the germanium detectors makes them better suited to constrain new physics in light mediator models. It should be noted however that Si HV detectors will have a lower threshold than their Ge counterparts and could play a very interesting complementary role should there be any sign of new physics.}.
The exposure corresponds to five years of operation with an 80\% live time and the background levels for electron and nuclear recoils in each detector configuration are taken from the predictions in Ref.~\cite{Agnese:2017jvy}.
We include an overall 75\% (85\%) signal efficiency for the iZIP (HV) detectors after analysis cuts \cite{Agnese:2016cpb}, though we neglect any energy dependence in the efficiency.
In our analysis, we compute separate bounds for electron and nuclear recoils using the iZIP configuration and a joint limit (adding electrons and nuclear recoils) for the HV detectors.
The nuclear-recoil threshold energies, both assumed to have been derived from a phonon signal, have been converted into electron-equivalent thresholds as per Eq.\,(8) of Ref.~\cite{Agnese:2017jvy} using the Lindhard model for the ionisation yield~\cite{osti_4701226}. The maxima of the energy windows resemble the approximate range of the detectors for the ER and NR+ER analyses.

\begin{table}[t]
\begin{center}
    \begin{tabular}{l c c c c c}
    \hline\\[-1ex]
        Experiment &  $\varepsilon$ (t$\cdot$yr) & NR (\si{\kilo\electronvolt}\nr{}) & ER (\si{\kilo\electronvolt}\ee{}) &  NR + ER (\si{\kilo\electronvolt}\nr{}) \\[1ex] 
        \hline \hline
        \\[0.05ex]
        G2-Ge\quad (SuperCDMS iZIP\cite{Agnese:2016cpb}) & 0.056 & [0.272, 10.4]  & [0.120, 50] & - \\ [1ex]
        \hspace*{6.8ex}\quad (SuperCDMS HV \cite{Agnese:2016cpb})&0.044& - & - & [0.040, 2]\\[1ex]
        G2-Xe\quad  (LZ \cite{Mount:2017qzi}) & 15 & [3, 5.8] & [2, 30] & [0.7, 100] \\[1ex]
        \hline \\
        G3-Xe\quad (DARWIN \cite{Aalbers:2016jon}) & 200 & [3, 5.8] & [2, 30] & [0.6, 100]\\[1ex] 
        G3-Ar\quad (DarkSide-20k \cite{Aalseth:2017fik}) & 100 &  - & [7, 50] & [0.6, 15]\\
        [1ex] 
        \hline
    \end{tabular}
\end{center}
\caption{\label{table:exps}
Simplified configurations for direct detection experiments used in this work. The nuclear recoil (NR) and electron recoil (ER) energy windows are given in nuclear recoil and electron-equivalent energies respectively. For the combined NR + ER analyses, the energies are given in nuclear recoil equivalent units.}
\end{table}

The G2-Xe and G3-Xe experiments have been inspired by the upcoming multi-ton liquid xenon (LXe) experiments LZ and DARWIN, respectively, and also serve as a proxy for competing experiments XenonNT \cite{Aprile:2015uzo} and PandaX \cite{Cao:2014jsa} using the same target. Both of their target nuclear-recoil thresholds are above the maximum of the solar neutrino spectra; they have been designed as such to minimise this signal, which is a background for WIMP searches. However, the LUX collaboration has been able to set thresholds as low as \SI{1.1}{\kilo\electronvolt}\nr{}, with a \SI{3.3}{\kilo\electronvolt}\nr{} energy threshold at 50\% detector efficiency in both the S1 and S2 channels required for ER/NR discrimination \cite{Akerib:2015rjg}.  In light of this and the advent of new analysis techniques allowing for even lower energy thresholds \cite{Akerib:2019zrt}, we have set thresholds for both LZ and DARWIN to be similar to this larger result. Above this threshold we can perform a nuclear recoil analysis with 99.5\% rejection of ER backgrounds with a 50\% acceptance of NR signal events \cite{Mount:2017qzi, Baudis:2013qla}. We take our backgrounds from Ref. \cite{Mount:2017qzi} for LZ and Ref. \cite{Baudis:2013qla} for DARWIN. In both cases, we extend our analyses beyond the energy range for which backgrounds are specified, taking the conservative assumption of a flat background spectrum within our energy range \cite{Akerib:2015cja}.
The lowest energy measurements in LXe to date have succeeded in detecting an ionisation (S2) signal at energies equivalent to \SI{0.3}{\kilo\electronvolt}\nr{} for nuclear recoils \cite{Lenardo:2019fcn} and \SI{0.186}{\kilo\electronvolt}\ee{} for electron recoils. Abandoning NR/ER discrimination which requires an S1 scintillation signal, we have set lower thresholds at approximately twice this former value for the G3-Xe setup and at \SI{0.7}{\kilo\electronvolt}\nr{} for the G2-Xe configuration - the S2-only threshold achieved by XENON100 \cite{Aprile:2016wwo}. We have conservatively assumed that the thresholds for these experiments for the  NR+ER analyses are equal in value in both \si{\kilo\electronvolt}\nr{} and \si{\kilo\electronvolt}\ee{} to avoid extrapolating the Lindhard model to energies at which it has not been tested experimentally\footnote{We have explicitly checked that extrapolating the Lindhard model to low energies would yield at most a 30\% improvement to our limits in the very low mass plateau, where $M_{A'} \ll 10^{-4}$ GeV, and a negligible difference in the region of interest for our work (i.e.~for $M_{A'} \gtrsim 10^{-3}$ GeV).}. The maximum recoil energy has been set at \SI{30}{\kilo\electronvolt}\ee{} ($\sim\SI{100}{\kilo\electronvolt}\nr{}$ using the Lindhard model) - the energy at which the double-$\beta$ decay of $^{136}$Xe, a considerable background for all LXe detectors, is expected to dominate over the solar neutrino signal  \cite{Mount:2017qzi}. 

Finally, the G3-Ar configuration has been based on the future DarkSide-20k detector, for which we also consider a five-year running time. The energy threshold values required for signal discrimination were found to be too high to give competitive constraints from the nuclear recoil spectrum \cite{Agnes:2018ves}, so only the ER-analysis was performed at this threshold. Since the only benefit of taking this higher threshold would be to minimise the background on an NR analysis, we have also taken the much lower threshold required for an ionisation (S2) signal with no ER/NR discrimination of \SI{0.6}{\kilo\electronvolt}\nr{}, as achieved by the DarkSide-50 experiment \cite{Agnes:2018ves}. The maximum of the energy window have been taken from the experimental design report \cite{Aalseth:2017fik} for the ER analysis and placed at \SI{15}{\kilo\electronvolt}\nr{} for the NR+ER analysis. The backgrounds we take from Fig. 7 of Ref. \cite{Agnes:2018ves}. As these do not extend to our maximum energy, we assume that the spectrum is flat above \SI{15}{\kilo\electronvolt}\nr{}, which we justify by comparison with Fig. 3 in the same paper.

\begin{figure}
\includegraphics[width=\textwidth]{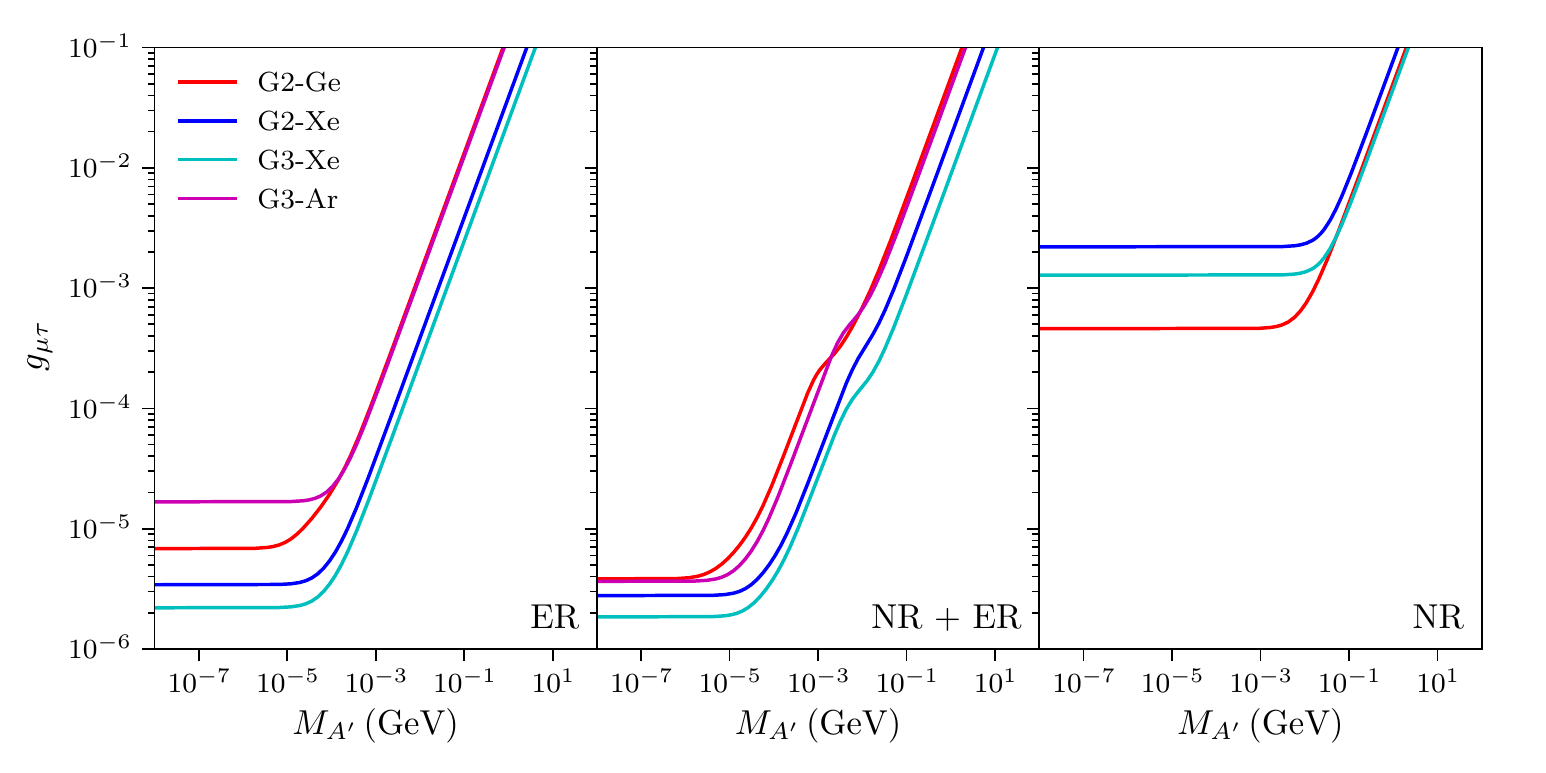}
\caption{Exclusion limits derived at 90\% confidence level  for (left) an ER, (middle) an NR + ER,  and (right) an NR analysis for the experiments considered in this work, assuming a high metallicity SSM.}
\label{fig:g_lims}
\end{figure}

For each of these setups, we have derived a 90\% C.L.~exclusion line assuming that no signal for new physics is observed. For a fixed value of the mediator mass, $M_{A'}$, we determine the value of the coupling, $g_{\mu \tau}$, for which 90\% of hypothetical experiments would expect to see an excess over the SM prediction. We have assumed that $N$ follows a Poisson distribution and varied the mean number of events to calculate this. The resulting exclusion lines for each experimental configuration are shown in Fig.\,\ref{fig:g_lims}. 
We separate the results for electron and nuclear recoils, and confirm (as previously pointed out in the literature \cite{Cerdeno:2016sfi}) that electron recoils  are  optimal to test very light mediator masses, whereas nuclear recoils excel for masses above $m_{A^\prime}>0.01$~GeV.\footnote{The sensitivity of direct detection experiments to new physics in the neutrino sector can also be interpreted as an increase in the so-called neutrino floor, below which neutrino-nucleus coherent scattering becomes a dominant background for dark matter searches \cite{Boehm:2018sux,Sadhukhan:2020etu}.} 
As we will see in the next Section, in the \Umt scenario, experimental constraints rule out the regions with very light mediator masses (in the coupling range suitable to explain $(g-2)_\mu$). Therefore, the best limits derived from direct detection are mostly based on the NR (or NR+ER) analysis. 
In this regime, the most advantageous strategy is to increase the experimental exposure, rather than further decreasing the threshold. This favours large experiments (such as LZ or DARWIN) versus smaller low-threshold detectors.

\section{Discussion}
\label{sec:discussion}

%
%
\begin{figure}[t]
\begin{center}
\includegraphics[width=\textwidth]{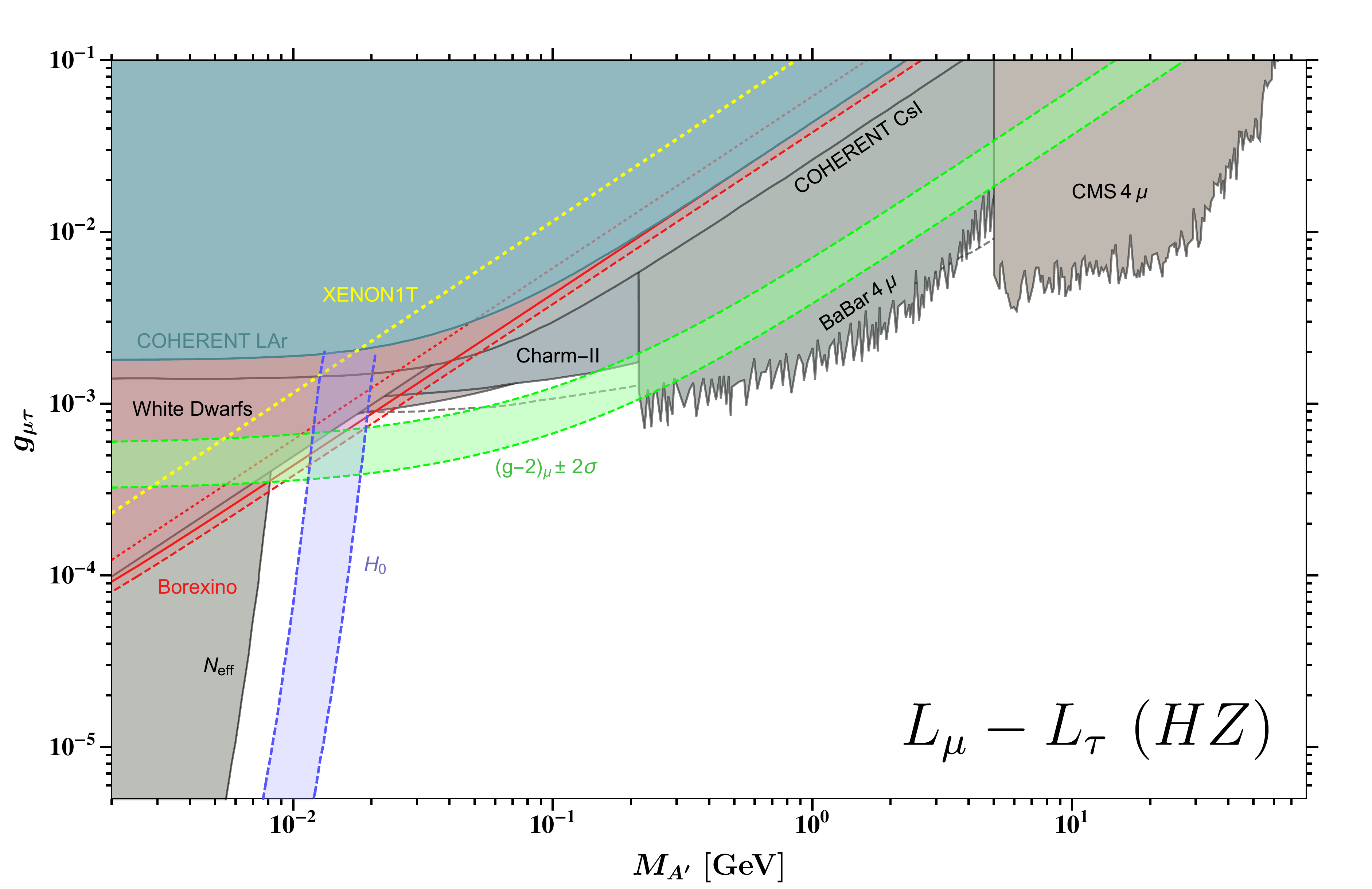}
\includegraphics[width=\textwidth]{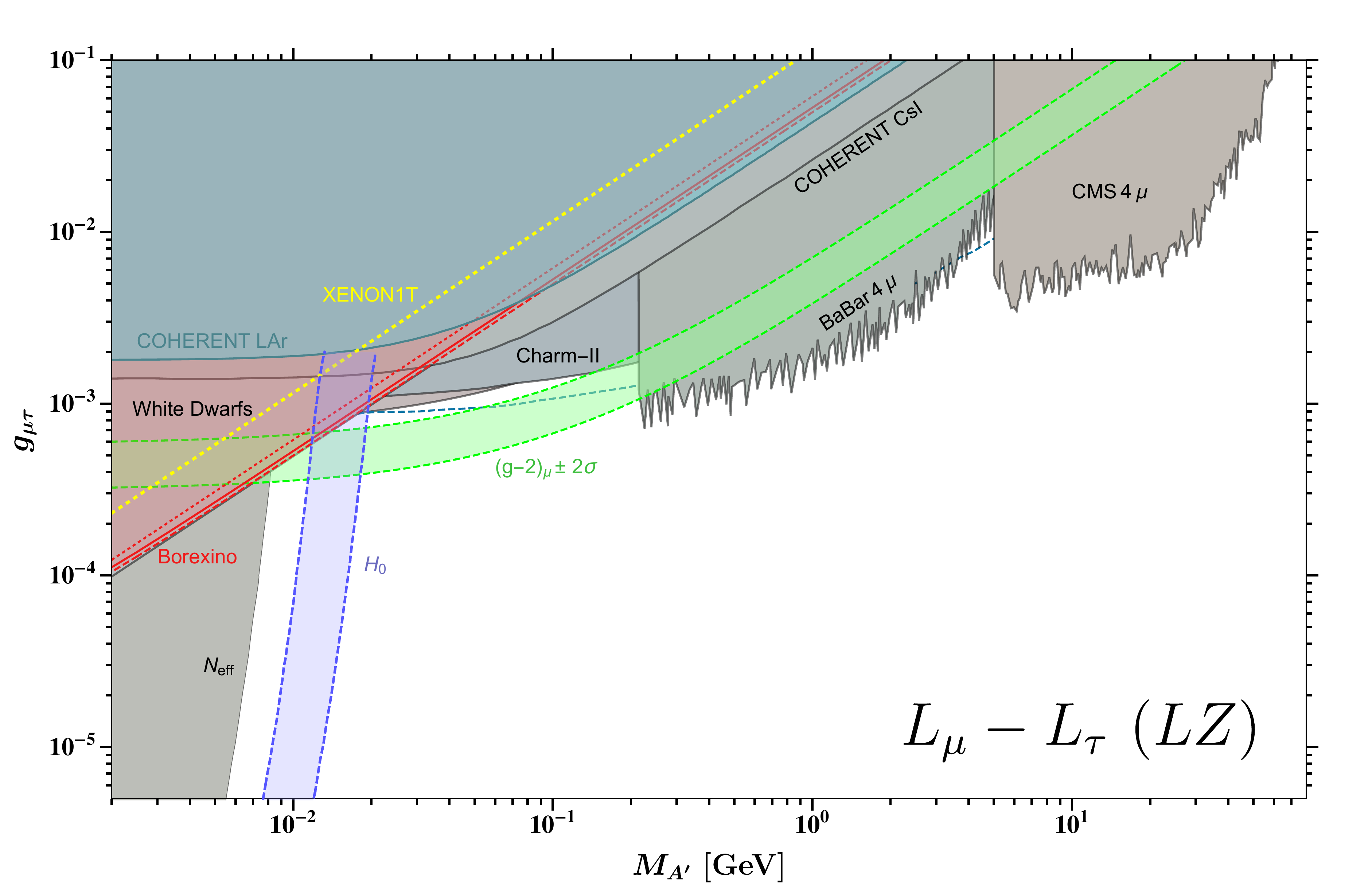}
\end{center}
\caption{\label{fig:curr_lims} Current constraints on the parameter space of the gauge boson of a minimal \Umt gauge group in grey. The green and blue bands show the regions favoured by $(g-2)_\mu$ and $H_0$, respectively. The yellow dotted line is the upper limit derived form the recent XENON1T result~\cite{Aprile:2020tmw}. Top: The red area corresponds to the  Borexino exclusion limit in the scenario of a high metallicity (HZ) Sun. Bottom: Same but for a low metallicity (LZ) Sun. See text in~\cref{sec:discussion} for detailed explanations.
}
\end{figure}
%
%

After having detailed in~\cref{sec:constraints} how neutrino-electron and -nucleus scattering can be used at various different experiments to constrain models of gauged \Umt, we want to discuss our results in the following section. \par

We show existing bounds on the \Umt gauge boson from white dwarf cooling, neutrino trident production, coherent scattering and four-muon searches in grey in~\cref{fig:curr_lims}. 
Concerning neutrino trident production, the most stringent bound is in principle due to the CCFR result~\cite{Altmannshofer:2014pba}. However, it has been pointed out in~\cite{Krnjaic:2019rsv} that the corresponding analysis did not take into account a background from diffractive charm production and hence needs further review. Thus, we show the corresponding limit obtained by CCFR only as a dashed black line in~\cref{fig:curr_lims} until this issue is resolved.
The region favoured by the observed deviation in $(g-2)_\mu$ at the $2\,\sigma$ level is illustrated by the green band. The region in parameter space explaining the tension in $H_0$, corresponding to $\Delta N_\text{eff} \sim 0.2 - 0.5$~\cite{Escudero:2019gzq}, is shown as the blue band\footnote{The  increased value of $N_\text{eff}$  only worsens the corresponding tension in $\sigma_8$ very mildly~\cite{Aghanim:2018eyx,Hildebrandt:2018yau}.\\}. Furthermore, we take the contour of $N_\text{eff}=4$ derived in~\cite{Escudero:2019gzq} as the conservative limit where a model of gauged \Umt should be firmly ruled out by cosmological observations. We display our limit derived from the liquid argon run of COHERENT outlined in~\cref{sec:constraints_coherent} as the cyan area. It can be seen that it has a similar sensitivity than the previous caesium-iodide run, but is slightly less constraining.
The yellow line represents the limit derived from the recent electron recoil data from XENON1T~\cite{Aprile:2020tmw}. As discussed in~\cref{sec:xenon1t} the simplistic treatment of the background in the chi squared test performed to derive the limit leads to an overly optimistic constraint. This limit lies well within previously excluded parameter space, and a more careful treatment of the background would only lead to a weaker bound. We therefore conclude that the current XENON1T constraint is subdominant with respect to the other existing bounds in the \Umt parameter space.
\par
In the top panel of~\cref{fig:curr_lims} we show our reevaluated limits from Borexino in red for the case of a high metallicity Sun. The solid red line corresponds to the limit derived with the more recent 2017 Borexino data set~\cite{Agostini:2017ixy}. The dashed line corresponds to the updated limit using the older 2011 data set~\cite{Bellini:2011rx}, including a more complete treatment of the theoretical and experimental uncertainties. We can see that the limit derived from the more recent data set is less constraining, due to a difference in the two measurements of the scattering rate which is explainable by a statistical fluctuation. Hence, we take the band between the solid and dashed line as the approximate envelope of the true  Borexino limit, which should be derived with a dedicated analysis combining the two results  in a single chi-squared analysis with appropriate handling of any shared systematic uncertainties. Nevertheless, even in the more conservative case the updated Borexino limit is significantly more constraining than the previously derived bound in~\cite{Bilmis:2015lja}, which is shown by the red dotted line. In particular, the updated limit rules out previously untested parameter space relevant for the explanation of $(g-2)_\mu$ above
\begin{equation}
    \frac{g_{\mu \tau}}{M_{A'}} \gtrsim \frac{0.043}{\text{GeV}} \qquad \qquad (HZ)\,,
\end{equation}
and hence yields the most stringent low-mass bound on this region of parameter space.
Most noticeably, it already excludes part of the region where the $(g-2)_\mu$ and $H_0$ bands overlap and a simultaneous explanation of the two would be in principle possible. 

%
%
\begin{figure}[t]
\begin{center}
\vspace{-1cm}
\includegraphics[width=\textwidth]{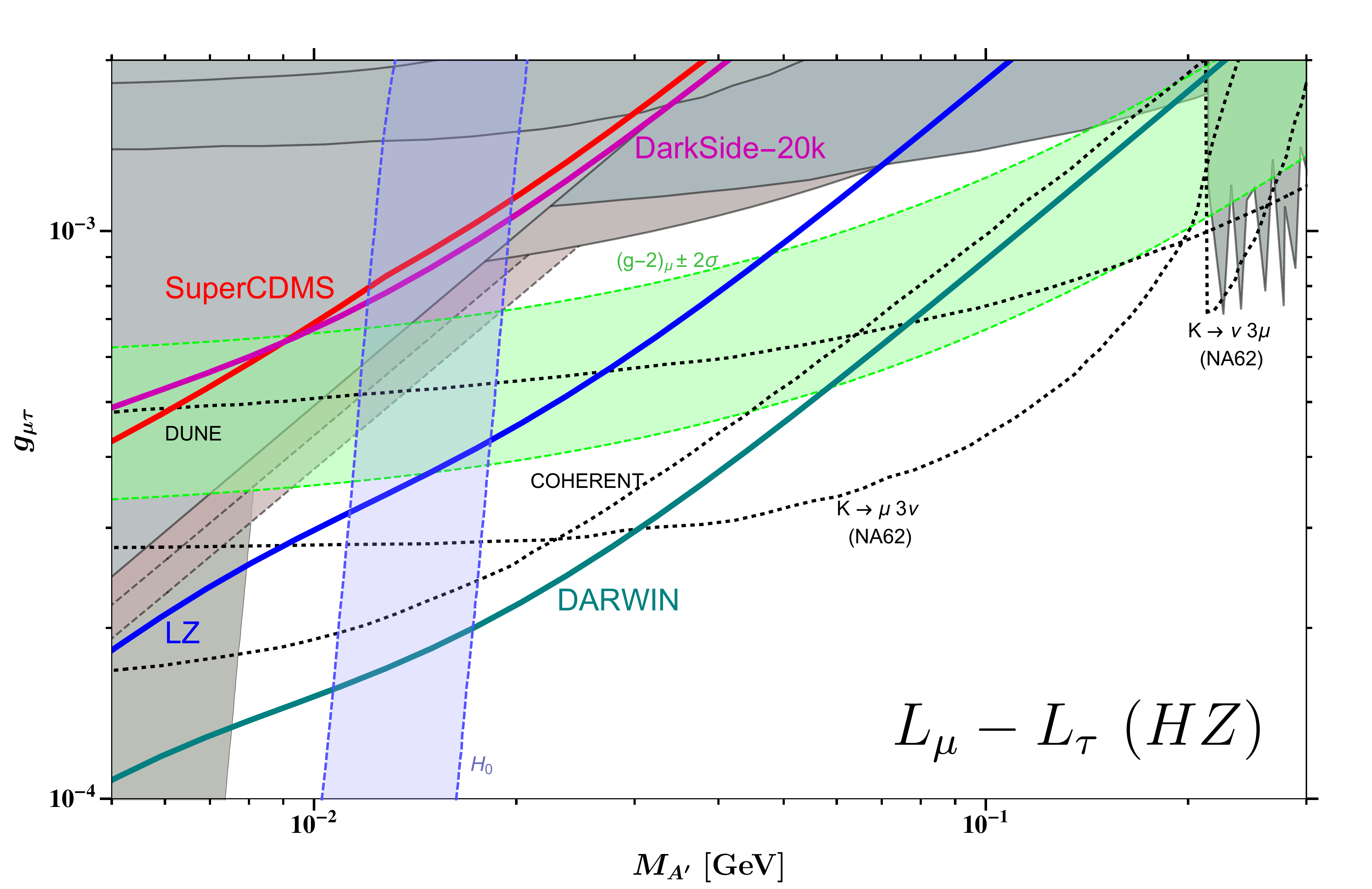}
\includegraphics[width=\textwidth]{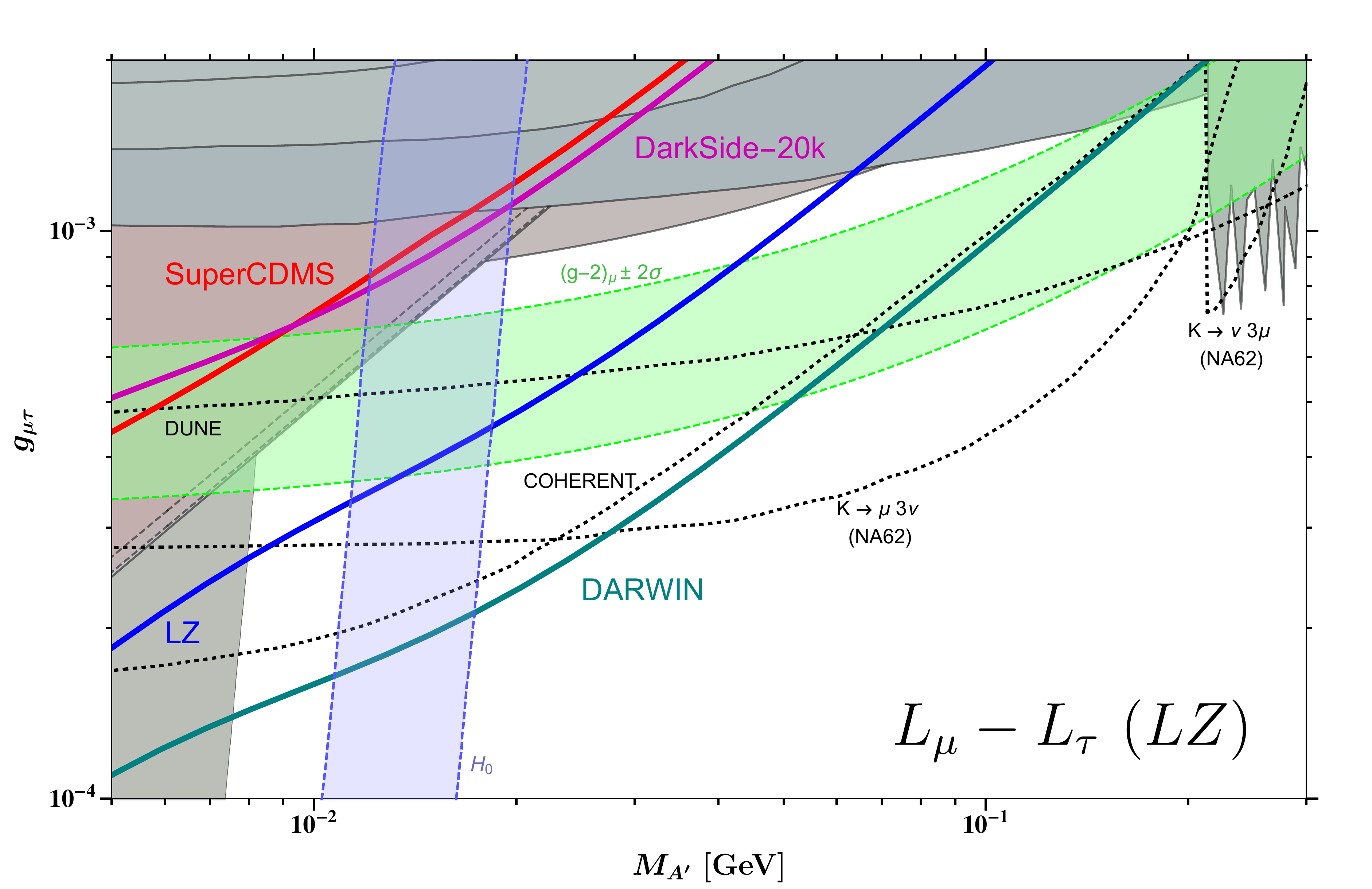}
\end{center}
\caption{\label{fig:proj_lims} 
Projected sensitivities of the direct detection experiments SuperCDMS, LZ, DARWIN and DarkSide-20k illustrated by the red, blue, turquoise and pink lines, respectively,  in the region of parameter space of a \Umt gauge boson relevant for an explanation of $(g-2)_\mu$ and $H_0$ (represented by the green and blue bands). Existing constraints are shown in  grey in the background. For comparison we show projections for COHERENT, DUNE and NA62 by the black dotted lines. Top: Sensitivities obtained with solar neutrino fluxes in the case of a high metallicity (HZ) Sun. Bottom: Same but for a low metallicity (LZ) Sun. See text in~\cref{sec:discussion} for detailed explanations.  
}
\end{figure}
%
%

\par
The Borexino limits for the case that the Sun is a low metallicity star are shown in the lower panel of~\cref{fig:curr_lims}. The general picture is quite similar to the high metallicity case. However, as the neutrino fluxes are generally lower in the low metallicity case the corresponding  SSM  prediction is also lower in the low metallicity case. Overall this leads to less constraining bounds from Borexino, which in the most conservative case rule out \Umt gauge bosons with
\begin{equation}
    \frac{g_{\mu \tau}}{M_{A'}} \gtrsim \frac{0.053}{\text{GeV}} \qquad \qquad (LZ)\,.
\end{equation}
Nevertheless, Borexino yields constraints which are competitive with those from white dwarf cooling, the most stringent previous bounds on the region of parameter space where $(g-2)_\mu$ is still allowed.
Most interestingly, the Borexino bounds still allow for a significantly larger part of the overlap region of the $(g-2)_\mu$ and $H_0$ bands in the low metallicity case. The  intriguing fact that a simultaneous explanation of the two tensions is still allowed in both the high and low metallicity scenario makes this region where the two bands overlap a prime target for future experimental searches. 
\bigskip

Motivated by the high sensitivity of solar neutrino scattering at Borexino  we next want to discuss future sensitivities of various experiments to probe the relevant region in parameter space. In~\cref{fig:proj_lims} we show projections for various different experiments on the relevant parameter space. The coloured solid lines show the envelope (i.e.~the most constraining bound for a given mass $M_{A'}$) of the projected limits obtained from analysing electron recoils (ER), nuclear recoils (NR) and the combination of the two (NR+ER) (cf.~\cref{fig:g_lims} for the individual limits) for the various dark matter direct detection experiments discussed in~\cref{sec:future_dm}. The results for SuperCDMS, LZ, DARWIN and DarkSide-20k are illustrated by the red, blue, turquoise and pink lines, respectively. For comparison we show previously derived projections from kaon decays at NA62~\cite{Krnjaic:2019rsv}, neutrino trident production at DUNE~\cite{Altmannshofer:2019zhy,Ballett:2019xoj} and a future $10$~t\,$\cdot$\,yr run of COHERENT with a NaI/Ar target~\cite{Abdullah:2018ykz} as black dotted lines.
\par
The projections for the direct detection experiments in the top panel of~\cref{fig:proj_lims} were derived using solar neutrino fluxes for the case of a high metallicity Sun. To begin with, it can be seen that SuperCDMS will not be able to improve over the updated Borexino limits. The envelope for SuperCDMS in the shown region is mainly due to the HV analysis not discriminating between electron and nuclear recoils and the iZIP analysis of nuclear recoils only. We explicitly checked and found that the major limitation of SuperCDMS is a nuclear background of $^{206}$Pb decays. Trying to reduce this background further would have the most effect on the limit and potentially allow SuperCDMS to gain sensitivity of unprobed parameter space.  
Next, we notice that  DarkSide-20k is very similar in reach compared to SuperCDMS. The achieved sensitivity is entirely due to the combined NR+ER analysis, as in the setup where discrimination of ER versus NR is possible, the corresponding threshold is too high to see solar neutrino scattering events through nuclear recoils.
In contrast to this, we observe that LZ and, most prominently, DARWIN will achieve the best sensitivity to such solar neutrino scattering events and can probe large parts of the region of parameter space relevant for explaining $(g-2)_\mu$. In particular, both experiments can probe the region where the preferred bands of $(g-2)_\mu$ and $H_0$ overlap. These are  especially promising prospects in the case of LZ as this is a G2 experiment which will happen in the near future. In this context, we want to note that this interesting region can also be probed entirely in $K\to \mu 3\nu$ at NA62 and in a future COHERENT run, while neutrino trident production at DUNE still has some sensitivity.
\par
In the lower panel of~\cref{fig:proj_lims}  we show the corresponding projections for the direct detection experiments derived with the neutrino fluxes from a low metallicity Sun. Overall the sensitivities are slightly worse due the generally lower flux of $\mathrm{^7Be}$ and $\mathrm{^8B}$ neutrinos. Nevertheless, the general picture does not differ substantially from the high metallicity case. 
The most noticeable difference is that due to the weaker Borexino limits, a large part of the   parameter space relevant for a simultaneous explanation of $(g-2)_\mu$ and $H_0$ is still viable.

The results of \cref{fig:proj_lims} clearly show that the optimal search strategy to cover the $(g-2)_\mu$ solution in the \Umt model with direct detection experiments is to look for coherent neutrino-nucleus scattering and push for large exposures. For this range of mediator masses, electron recoils are a subdominant contribution, mostly due to the related background. Also, lowering the energy threshold is only effective for probing small mediator masses with nuclear recoils (see the rightmost panel in~\cref{fig:g_lims}). For this reason, xenon-based experiments as LZ outperform low-threshold detectors such as SuperCDMS.

\section{Conclusions}
\label{sec:conclusions}

 Models of gauged \Umt symmetry have received a lot of attention in recent years since their special flavour coupling structure can help to explain tensions in the observed value of the anomalous magnetic moment of the muon $(g-2)_\mu$ and of the Hubble parameter $H_0$. Motivated by these intriguing prospects, in this article we have studied the potential sensitivity of solar neutrino scattering to probe these conjectured explanations within \Umt. 
 We have summarised our main results in~\cref{fig:curr_lims,fig:proj_lims}. \par
 \begin{itemize}
     \item We have reevaluated previous results from solar neutrino-electron scattering at Borexino. We find that these have been previously underestimated (see~\cref{sec:b-l} for a similar result for $U(1)_{B-L}$) and that they yield the most constraining current limit in the region where $(g-2)_\mu$ can be explained. 
     In the scenario of a high metallicity Sun, which is currently favoured by experimental data, the updated limits exclude a substantial part of previously untested parameter space relevant for a simultaneous explanation of $(g-2)_\mu$ and $H_0$.
     In the case of a low metallicity Sun, Borexino cannot exclude large parts of the simultaneous explanation of $(g-2)_\mu$ and $H_0$, but still yields constraints which are competitive with previous leading bounds.
     
     \item We have analysed recent data of the liquid argon run of the COHERENT experiment and derived the corresponding limits on the parameter space of a \Umt boson. While these limits are similar in reach to those derived from a previous caesium-iodide run they are slightly less constraining and cannot exclude any untested parameter space.   
     
     \item In view of the recently reported excess of electron recoil events in XENON1T, we have shown that a \Umt boson cannot account for the observed excess. Instead, we have used the corresponding data to derive a constraint on the parameter space of \Umt. We have found that this does not yield a competitive bound compared with already existing limits.
     
     \item Furthermore, we have shown that upcoming and future dark matter direct detection experiments are a very promising means of probing the phenomenologically interesting region of parameter space where a simultaneous explanation of $(g-2)_\mu$ and $H_0$ is possible. 
     The main search strategy to cover this region is to look for coherent neutrino nucleus scattering with increasing exposures, while electron recoils and lower thresholds are not as optimal.
     Thus, while SuperCDMS and DarkSide-20k will not be able to set competitive limits without major improvements of the detectors, LZ and DARWIN will be able to test a significant part of the parameter space relevant for $(g-2)_\mu$. Considering that LZ is a second generation DM experiment and will happen in the near future these are very exciting prospects. 
 \end{itemize}

 We eagerly await the upcoming results from the measurement of the muon anomalous magnetic moment at the Fermilab E989 experiment. Should there be a confirmation ($5\,\sigma$ discovery) of the excess over the SM value, our work suggests that direct detection experiments would play a leading role in identifying the new physics model by probing the \Umt parameter space in complementarity with COHERENT, Dune and NA62.

 \par\bigskip
 {\it Note:} Upon completion of this work we became aware of a very recent article~\cite{Sadhukhan:2020etu} complementary to our analysis that explored  modifications of the \coherent neutrino floor in a gauged \Umt.

\section*{Acknowledgements}
We thank M.~Bauer, A.~Biek\"otter, M.~Escudero, J.~Jaeckel, P.A.N.~Machado, D.K.~Papoulias, R.~Tayloe and M.~T\'ortola for helpful discussions during the preparation of the manuscript.
DGC acknowledges financial support from the project SI2/PBG/2020-00005 and is also supported in part by the Spanish Agencia Estatal de Investigaci\'on through the grants PGC2018-095161-B-I00 and IFT Centro de Excelencia Severo Ochoa SEV-2016-0597, and the Spanish Consolider MultiDark FPA2017-90566-REDC.
PF is funded by the UK Science and Technology Facilities Council (STFC) under grant ST/P001246/1.


\appendix 
\section{Loop-induced kinetic mixing}
\label{sec:loopmix}

In order to determine the effects of a new neutral vector boson on neutrino scattering it is crucial to correctly incorporate the interference term between the new boson and the SM $Z$. In the case of \Umt the sign of the interference critically depends on the sign of the kinetic mixing term relative to the gauge interactions. Hence, in the following we will carefully derive the loop-induced kinetic to unambiguously determine its sign. \par
In order to maintain full generality, we consider a set of (chiral) fermions $\psi_i$ carrying charges $Q_a^i$ and $Q_b^i$ under two Abelian gauge groups $U(1)_a\times U(1)_b$ according to
\begin{equation}
    \C{L} = -\frac{1}{4} X_a^{\mu\nu} X_{a\mu\nu} -\frac{1}{4} X_b^{\mu\nu} X_{b\mu\nu} + i \bar \psi_i \slashed D \psi_i \,,
\end{equation}
with $D_\mu=\partial_\mu + i\, g_a\, Q_a^i\, X_{a\mu} + i\,g_b\, Q_b^i\, X_{b\mu}$. At the loop level the fermions $\psi_i$ will induce a vacuum polarisation term for the two bosons $X_a$ and $X_b$ via the diagram
\begin{align}
    \vcenter{\hbox{  \adjincludegraphics[ trim={0 0 {.1\width} {.08\height}},clip, width=0.35\textwidth]{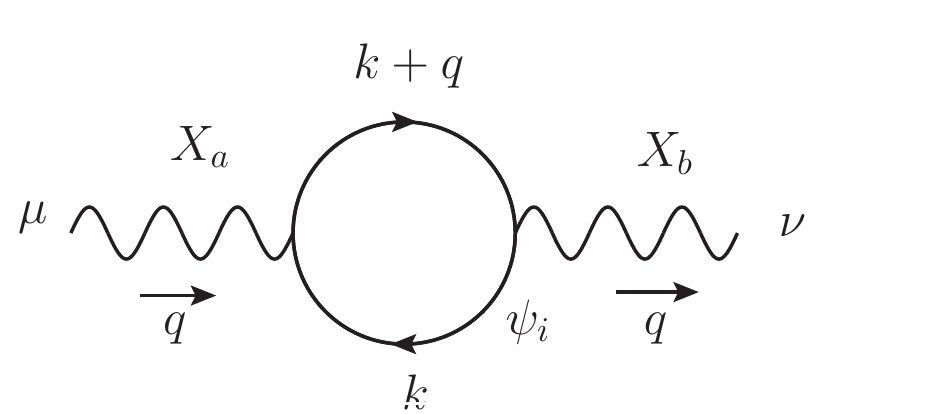}}} &= \quad i\, \Pi^{\mu\nu}_{ab}(q)\,,
\end{align}
\newline
which is analogous to the vacuum polarisation of the photon in QED. Summing over all the fermion fields $\psi_i$ running in the loop, the amplitude of the vacuum polarisation diagram is given by
\begin{equation}
    i\, \Pi^{\mu\nu}_{ab}(q) = \sum_i (-1)(-i g_a Q^i_a)(-i g_b Q^i_b)\int \frac{d^4k}{(2\pi)^4}
    \,\mathrm{tr}\left[ \gamma^\mu \frac{i(\slashed k + m_i)}{k^2 - m_i^2}\gamma^\mu \frac{i(\slashed k +\slashed q + m_i)}{(k+q)^2 - m_i^2}\right]\,.
\end{equation}
In order to evaluate the integral over the loop momentum we follow standard textbook methods\footnote{For a detailed treatment of the analogous calculation of the QED vacuum polarisation see e.~g.~\cite{Peskin:1995ev,Srednicki:2007qs}.} and introduce Feynman parameters to bring the integral into the form 
\begin{equation}
    i\, \Pi^{\mu\nu}_{ab}(q) = -4\, g_a g_b\sum_i Q^i_a Q^i_b\int_0^1 dx\int \frac{d^4k}{(2\pi)^4}
    \,\frac{k^\mu(k+q)^\nu + k^\nu(k+q)^\mu - g^{\mu\nu}(k\cdot(k+q)-m_i^2)}{\left[k^2 - m_i^2 + 2 x \, k\cdot q + x q^2\right]^2}\,.
\end{equation}
Completing the square and shifting the loop momentum to $\ell=k + xq$ we will finally end up with the expression for the vacuum polarisation as
\begin{equation}
    i\, \Pi^{\mu\nu}_{ab}(q) = -4\, g_a g_b\sum_i Q^i_a Q^i_b\int_0^1 dx\int \frac{d^4\ell}{(2\pi)^4}
    \,\frac{2\ell^\mu\ell^\nu - 2x(1-x)q^\mu q^\nu - g^{\mu\nu}(\ell^2-x(1-x)q^2-m_i^2)}{\left[\ell^2 - m_i^2 +x(1-x)q^2\right]^2}\,.
\end{equation}
The momentum integral is clearly divergent ($\sim \int d^4\ell(1/\ell^2)$) and we have to regularise it. We will use dimensional regularisation in $d=4-\varepsilon$ dimensions to compute the divergent integrals. The result of this calculation is given by
\begin{equation}
    i\, \Pi^{\mu\nu}_{ab}(q) = -i \, \left[g^{\mu\nu}q^2 - q^\mu q^\nu\right]\ \frac{g_a g_b}{2 \pi^2}\sum_i Q^i_a Q^i_b\int_0^1 dx \ x(1-x) \left\{ \frac{2}{\varepsilon}+\log\left(\frac{\mu^2}{\Delta_i} \right) + \C{O}(\varepsilon)\right\} \,,
\end{equation}
where $\mu$ denotes an arbitrary mass scale and $\Delta_i = m^2_i - x(1-x)q^2$. In an $\overline{\text{MS}}$ renormalisation scheme we will split off the divergent $2/\varepsilon$ piece and absorb it into a counterterm~\cite{Pich:1998xt}. Extracting the Lorentz tensor structure
\begin{equation}
     i\, \Pi^{\mu\nu}_{ab}(q) = i\left[g^{\mu\nu}q^2 - q^\mu q^\nu\right]\, \Pi_{ab}(q^2)\,,
\end{equation}
the finite renormalised part of the vacuum polarisation is then obtained to be
\begin{equation}
    \Pi_{ab}(q^2) = -\frac{g_a g_b}{2 \pi^2}\sum_i Q^i_a Q^i_b\int_0^1 dx \ x(1-x)\,\log\left(\frac{\mu^2}{\Delta_i} \right)\,.
\end{equation}
\par
The effect of this vacuum polarisation term will manifest itself through an effective mixing of the two gauge bosons $X_a$ and $X_b$ via
\begin{equation}
    \C{L}_\text{eff} = - \frac{\epsilon_{ab}}{2} X_a^{\mu\nu}X_{b\mu\nu}\,.
\end{equation}
In order to see how this comes about we have to match the diagram of the vacuum polarisation with the one originating from the mixing,
\begin{align}
    \vcenter{\hbox{\includegraphics[width=.25\textwidth]{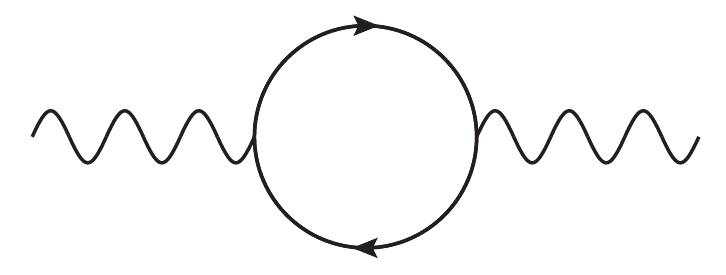}}} &\overset{!}{=} \vcenter{\hbox{\includegraphics[width=.25\textwidth]{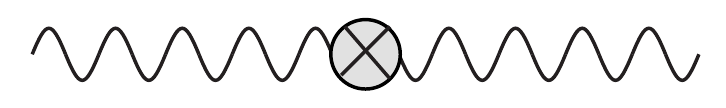}}} \\
    i\left[g^{\mu\nu}q^2 - q^\mu q^\nu\right]\, \Pi_{ab}(q^2) &= -i\left[g^{\mu\nu}q^2 - q^\mu q^\nu\right]\,\epsilon_{ab}\,.
\end{align}
This matching allows us to identify the effective mixing parameter as
\begin{equation}
    \epsilon_{ab} = - \Pi_{ab}(q^2) = \frac{g_a g_b}{2 \pi^2}\sum_i Q^i_a Q^i_b\int_0^1 dx \ x(1-x)\,\log\left(\frac{\mu^2}{\Delta_i} \right)\,.
\end{equation}

\section{Matter-induced Solar Neutrino Oscillations}
\label{sec:oscillations}

For a model with non-flavour-universal couplings to neutrinos, it is important to know the population of each neutrino flavour eigenstate, $\nu_\alpha$, incident on a potential experiment. To derive constraints from the scattering of solar neutrinos, we must therefore understand the matter-induced neutrino oscillations which take place inside the Sun.

Although solar neutrinos are produced through multiple processes, they are all generated in the electron neutrino eigenstate. The osciallations of the neutrinos as they propagate through the Sun are governed by the MSW effect \cite{Smirnov:2004zv}. The neutrinos propagate adiabatically through the solar medium with the relative abundances of the three mass eigenstates, $\nu_i$, fixed at the values they had immediately after production. However, as the neutrinos were produced in an environment with a large electron density, the mixing angles which relate the flavour eigenstates to the mass eigenstates in the solar core are modified from their in-vacuum values, leading to an energy-dependent suppression of the number of neutrinos reaching terrestrial experiments in the $\nu_e$ eigenstate. 

Once the neutrinos have left the Sun, they continue to propagate to the Earth in their mass eigenstates. As the distance from the Sun to the Earth is many orders of magnitude longer than the neutrino oscialltion length, the neutrinos separate into a decoherent mixture of mass eigenstates, so that the overall transition probability from the Sun's core to detection in an experiment is given by
\begin{equation}
P(\nu_e \rightarrow \nu_\alpha) = \sum_{i=1}^{3} |U_{\alpha i}|^2 P_i,
\end{equation}
with $P_i$ the probability of the neutrino arriving at the detector in the $\nu_i$ eigenstate and $U$ the PMNS matrix \cite{Giganti:2017fhf, Esteban:2018azc}.
Terrestrial matter effects could further modify the populations of neutrino eigenstates reaching the detector, leading to a day-night asymmetry as neutrinos arriving at night will have traversed the Earth \cite{Blennow:2003xw}. However, the effect is small at the low energies which are most relevant for this work, with an asymmetry of less than $10^{-3}$ for neutrinos below 1~MeV \cite{Aleshin:2013}. We therefore neglect it in this work.

So far, the only direct observations of solar neutrinos have been through interactions with electrons \cite{Agostini:2017cav,Aharmim:2011vm,Abe:2016nxk}. In the standard model, at the typical energies of solar neutrinos, it is valid to assume that the cross section for a muon- and tau-neutrino scattering with an electron is identical, as both proceed through the exchange of a $Z$-boson (the ``neutral current'' channel), whereas the scattering of electron-neutrinos has the additional ``charged current'' contribution from the exchange of a $W$-boson. Under this assumption, the only transition probability that must be known is the electron-neutrino survival probability, $P(\nu_e \rightarrow \nu_e)$. This quantity can be calculated with high precision in the two-neutrino approximation \cite{Balantekin:2003dc}, and this has been the subject of many previous works \cite{Nunokawa:2006ms,Berezinsky:2001uv,Vissani:2017wvk,Lopes:2013nfa}. However, in our model each flavour eigenstate has a different coupling, so it is important to know the relative abundance of each eigenstate.

As discussed in \cref{sec:nu-nsi}, previous works have shown that any non-flavour diagonal NSIs can affect the matter potential and can therefore change the neutrino oscillation probabilities \cite{Farzan:2017xzy}. However, as per \cref{eq:nsi_cancellation}, assuming that solar matter is electrically neutral means that the NSIs couplings generated in our model with protons and electrons exactly cancel and no net effect is expected on solar neutrino oscillations.

\begin{figure}[t]
\begin{center}
\includegraphics[width=0.7\textwidth]{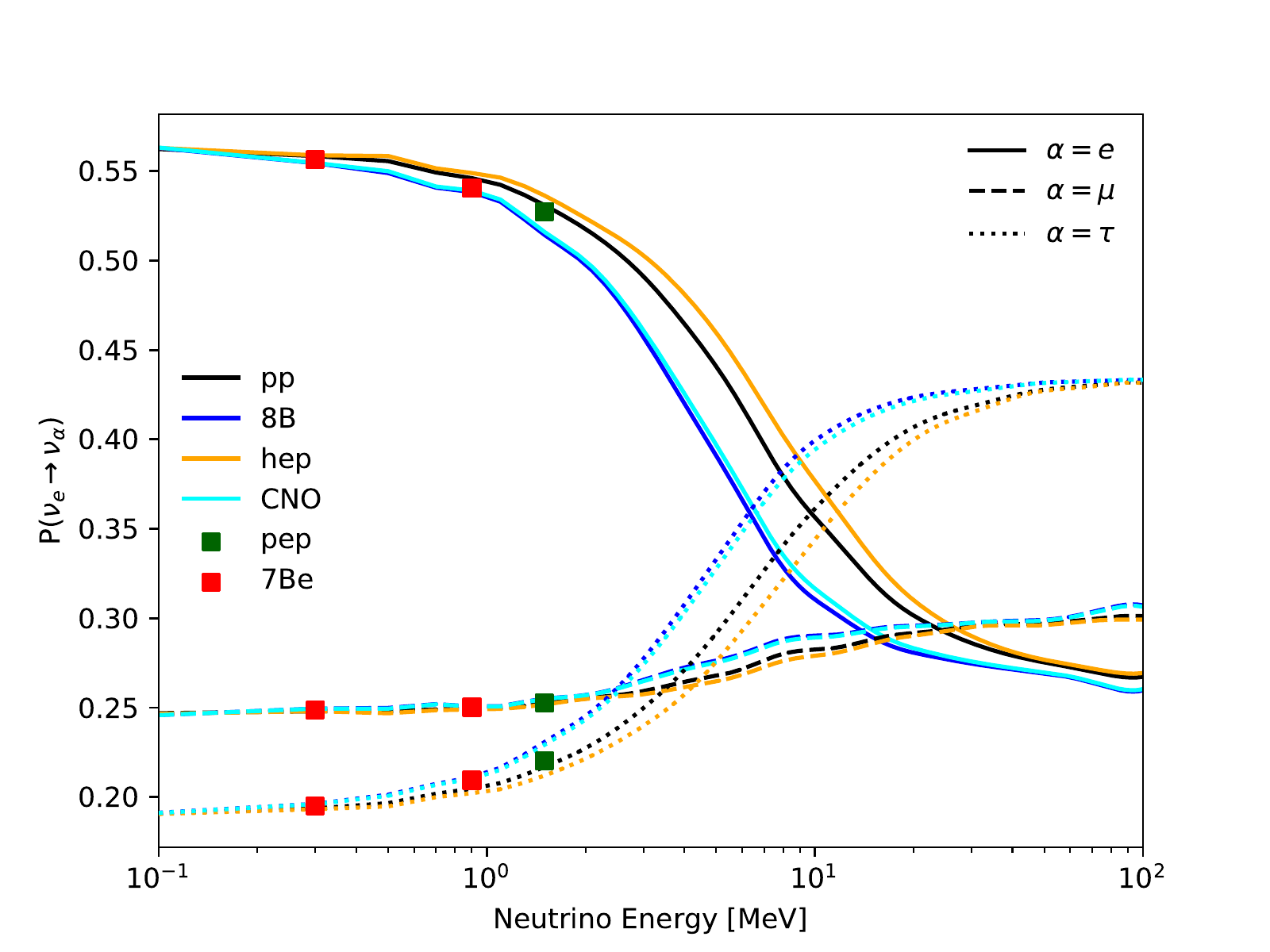}
\end{center}
\caption{\label{fig:nu_flavour_probs} Probability of finding a solar neutrino produced in the electron eigenstate in each of the $\alpha \in \{e, \mu, \tau \}$ flavour eigenstates after its journey to Earth. We plot the probabilities for each of the solar neutrino populations separately as they are produced in different regions of the Sun, and therefore have different oscillation rates. The CNO line is an average across the three populations of CNO neutrinos: the $\mathrm{^{13}N}$, $\mathrm{^{15}O}$, and $\mathrm{^{15}F}$ neutrinos.}
\end{figure}

Oscillation probabilities will also differ between the various solar neutrino fluxes, since they are produced in different regions of the Sun \cite{Lopes:2013nfa}. Fig. \ref{fig:nu_flavour_probs} shows the averaged three-flavour oscillation probabilities for all the relevant solar neutrino fluxes. The overall neutrino transition probabilities were calculated using nuSQuIDS \cite{Delgado:2014kpa}, a software package for numerically calculating neutrino oscillation probabilities in matter \cite{Casas:2016asi}.

\section{Updated analysis of Borexino data for $B-L$}
\label{sec:b-l}

%
%
\begin{figure}[t]
\begin{center}
\includegraphics[width=\textwidth]{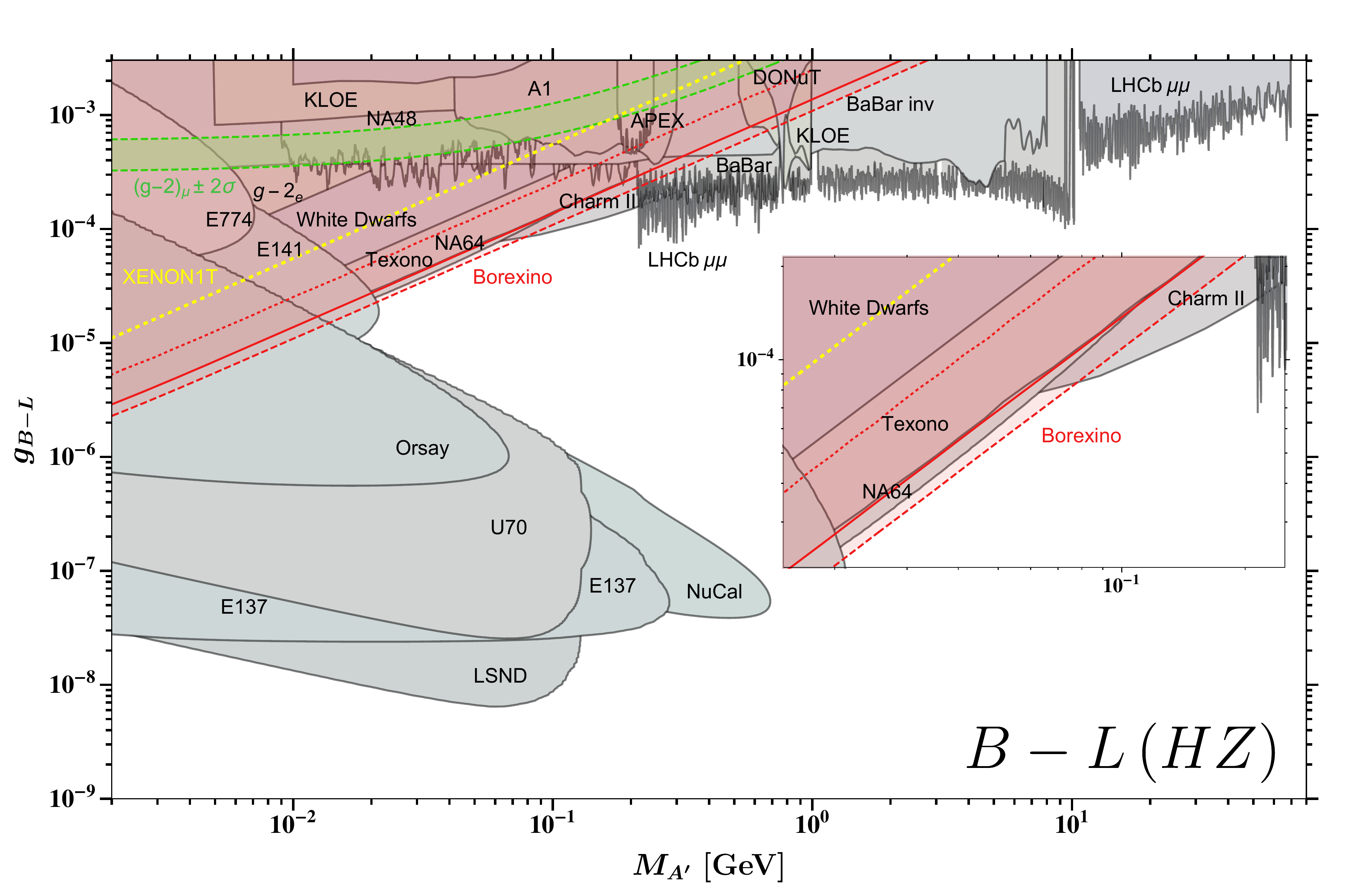}
\includegraphics[width=\textwidth]{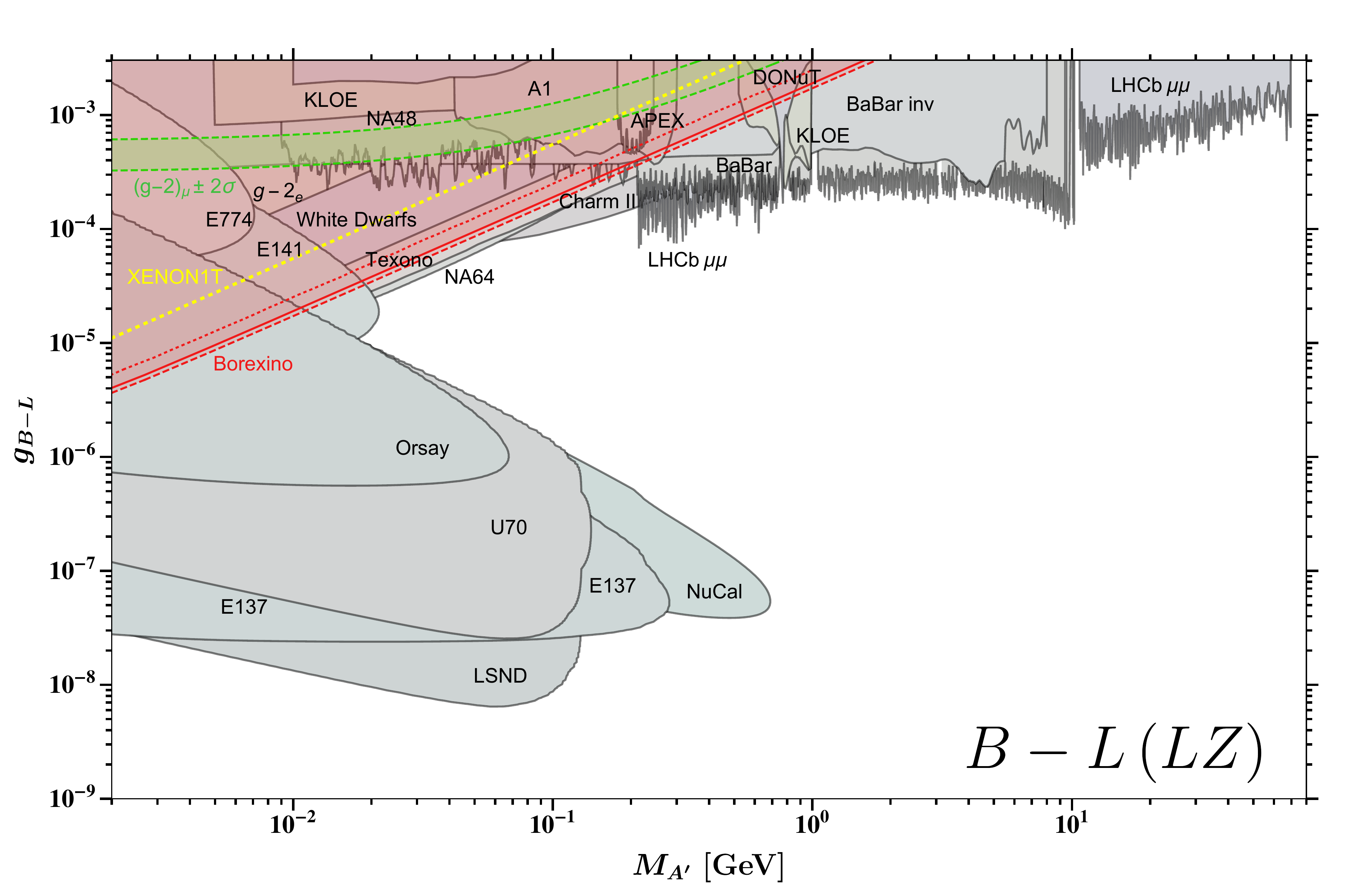}
\end{center}
\caption{\label{fig:bl_lims}
Current constraints on the parameter space of the gauge boson of a minimal $U(1)_{B-L}$ gauge group shown in grey. The green band shows the regions favoured by $(g-2)_\mu$. The yellow dotted line is the upper limit derived form the recent XENON1T result~\cite{Aprile:2020tmw}.  Top: The red area corresponds to the Borexino exclusion limit in the scenario of a high metallicity (HZ) Sun.  Bottom: As above, but with the Borexino limit calculated in the scenario of a low metallicity (LZ) Sun. See text in~\cref{sec:b-l} for detailed explanations.
}
\end{figure}
%
%

As mentioned in \cref{sec:constraints_borexino} neutrino-electron scattering at Borexino has been previously used to constrain models of gauged $U(1)_{B-L}$ in Ref.~\cite{Harnik:2012ni}. However, as was pointed out in Ref.~\cite{Bilmis:2015lja}, this first derivation neglected the interference term arising between the $U(1)_{B-L}$ gauge boson and the SM $Z$-boson. 
By virtue of~\cref{eq:sig_el} and the replacement of the NSI coupling induced by a $U(1)_{B-L}$ gauge boson,
\begin{equation}
       \varepsilon^{fP}_{\alpha\alpha} =  -\frac{g_{B-L}^2\, Q^{BL}_{f}}{2\sqrt{2}\, G_F\,(2 m_e E_R+M_{A'}^2)}\,,
\end{equation}
the differential neutrino-electron scattering cross section including interference is found to be,
\begin{align}\label{eq:sig_elec_bl}
     \frac{d \sigma_{\nu_\alpha\, e}}{d E_R} =   \frac{2\, G_F^2\,m_e}{\pi} & \ \Bigg\{ \,  \left[ {g^\alpha_1}^2+{g^\alpha_2}^2\left(1-\frac{E_R}{E_\nu}\right)^2 - g^\alpha_1\,g^\alpha_2 \frac{m_e \, E_R}{E_\nu^2} \right]  \notag \\
    &+ \frac{g_{B-L}^2}{\sqrt{2}\, G_F (2 m_e E_R + M_{A'}^2)} \left[ {g^\alpha_1}+{g^\alpha_2}\left(1-\frac{E_R}{E_\nu}\right)^2 - (g^\alpha_1+g^\alpha_2) \frac{m_e \, E_R}{2 E_\nu^2} \right]  \notag \\
    &+  \frac{g_{B-L}^4}{8\, G_F^2 (2 m_e E_R + M_{A'}^2)^2} \left[  1 + \left(1-\frac{E_R}{E_\nu}\right)^2 -  \frac{m_e \, E_R}{E_\nu^2} \right]\, \Bigg\} \,,
\end{align}
where the couplings $g^\alpha_1$ and $g^\alpha_2$ are given by~\cref{eq:cpls_esm}. While in the derivation of~\cite{Bilmis:2015lja} these interference effects have been taken into account, important  theoretical uncertainties on the SM and SSM prediction (including the uncertainty stemming from the solar metallicity problem) have been omitted. Combined with the consistent incorporation of the interference term in~\cref{eq:sig_elec_bl} the inclusion of these uncertainties  warrants for an updated analysis of the corresponding Borexino constraint on $U(1)_{B-L}$ as outlined in~\cref{sec:constraints_borexino}. 
\par
We show the results of our improved analysis in~\cref{fig:bl_lims}. Existing constraints on the parameter space of a $U(1)_{B-L}$ boson are shown as the grey areas\footnote{For detailed explanations of the various existing bounds as well as an overview of future projected sensitivities of a large number of experiments we refer to Ref.~\cite{Bauer:2018onh}.}.
In the top panel we show the reevaluated Borexino limit from the 2011 data set~\cite{Bellini:2011rx} as the red dashed line for the case of a high metallicity Sun. The constraint derived from the more recent 2017 data set~\cite{Agostini:2017ixy} is shown as the red solid line. Interpreting these two lines as an envelope of this true constraint, which would require the two results to be combined in a single chi-squared analysis with appropriate handling of any shared systematic uncertainties, we see that Borexino and NA64~\cite{NA64:2019imj} are setting the most stringent bound for moderate couplings in the range $M_{A'}\approx 20 - 80$ MeV. For comparison we show the previous bound derived in~\cite{Bilmis:2015lja} as the red dotted line. 
\par
We also show the results of our simplified analysis of the recent XENON1T result, discussed in detail in \cref{sec:xenon1t}. Although we have not performed the most complete analysis, we expect it to give an optimistic estimate of the limits, and since they do not cover any new regions of the parameter space we do not expect this result from XENON1T to supply competitive constraints on a $U(1)_{B-L}$ gauge boson within this mass region.
\par  In the bottom panel of~\cref{fig:bl_lims} we have evaluated the limits from Borexino and XENON1T in the case of a low metallicity Sun. In this case, the lower flux of $\mathrm{^7Be}$ neutrinos leads to a weaker constraint from Borexino than we obtain in the HZ case, and we do not exclude any more of the parameter space beyond the limits already set by NA64 ~\cite{NA64:2019imj}.
\par
Finally, we expect qualitatively quite similar results of an updated analysis of the Borexino data for other anomaly-free groups gauging a combination of $B$ and $L_e$, like $U(1)_{B-3L_e}$, $U(1)_{B-L_e-2L_\alpha}$ or $U(1)_{B-L_\alpha-2L_e}$ with $\alpha=\mu,\tau$.

\bibliographystyle{JHEP-cerdeno}
\bibliography{sample}

\end{document}